\documentclass[%
 reprint,
 amsmath,amssymb,
 aps,
 floatfix,
 prb
]{revtex4-2}
\usepackage{subfigure}
\usepackage{physics}
\usepackage{braket}
\usepackage{lipsum} 
\usepackage[colorlinks=true]{hyperref}
\usepackage{orcidlink}
\usepackage{graphicx}
\usepackage{dcolumn}
\usepackage{bm}\allowdisplaybreaks

\usepackage[normalem]{ulem}
\usepackage{dsfont}
\usepackage{amsmath}
\usepackage{amssymb}
\usepackage{mathrsfs}

\usepackage{bibentry}

\newcommand{\ad}[1]{\operatorname{ad}_{#1}}
\newcommand{\Ad}[1]{\operatorname{Ad}_{#1}}
\newcommand{\Prod}[2]{\left\langle #1, #2 \right\rangle}

\DeclareMathAlphabet{\mathcalligra}{T1}{pzc}{m}{it}

\begin{document}

\title{Fermi Surface Bosonization for Non-Fermi Liquids}

\author{SangEun Han\,\orcidlink{0000-0003-3141-1964}}
\thanks{These authors contributed equally to this work.}
\affiliation{Department of Physics, University of Toronto, Toronto, Ontario M5S 1A7, Canada}
\author{F\'{e}lix Desrochers\,\orcidlink{0000-0003-1211-901X}}
\thanks{These authors contributed equally to this work.}
\affiliation{Department of Physics, University of Toronto, Toronto, Ontario M5S 1A7, Canada}
\author{Yong Baek Kim\,\orcidlink{0000-0003-0473-8785}}
\affiliation{Department of Physics, University of Toronto, Toronto, Ontario M5S 1A7, Canada}

\date{\today}

\begin{abstract}
Understanding non-Fermi liquids in dimensions higher than one remains one of the most formidable challenges in modern condensed matter physics. These systems, characterized by an abundance of gapless degrees of freedom and the absence of well-defined quasiparticles, defy conventional analytical frameworks. Inspired by recent work [Delacretaz, Du, Mehta, and Son, \href{https://doi.org/10.1103/PhysRevResearch.4.033131}{Physical Review Research, 4, 033131 (2022)}], we present a procedure for bosonizing Fermi surfaces that does not rely on the existence of sharp excitation and is thus directly applicable to non-Fermi liquids. Our method involves parameterizing the generalized fermionic distribution function through a bosonic field that describes frequency-dependent local variations of the chemical potential in momentum space. We propose an effective action that produces the collisionless quantum Boltzmann equation as its equation of motion and can be used for any dimension and Fermi surface of interest. Even at the quadratic order, this action reproduces non-trivial results obtainable only through involved analysis with alternative means. By offering an alternative method directly applicable to studying the low-energy physics of Fermi and non-Fermi liquids, our work potentially stands as an important building block in advancing the comprehension of strange metals and associated phenomena.
\end{abstract}

\maketitle
\section{Introduction}

The description of metals is arguably the oldest problem in all of condensed matter physics~\cite{sommerfeld1928elektronentheorie,bloch1929quantenmechanik}. Despite this historical status, a satisfactory theoretical description of the low-energy physics of metals remains an outstanding problem made exceedingly hard by the continuum of gapless degrees of freedom located along a nontrivial submanifold of momentum space (i.e., the Fermi surface). When metals host well-defined quasiparticles, Landau's phenomenological Fermi liquid (FL) theory can be used as a bare-bones description to compute certain correlation functions at leading order. Nonetheless, in many cases of physical interest, interactions are so important that the system does not have well-defined quasiparticles while maintaining a sharp Fermi surface. These so-called non-Fermi liquids (NFL) are ubiquitous in the phase diagrams of strongly correlated systems. They can arise when metals are subjected to singular interactions mediated by critical bosonic modes~\cite{altshuler1994low, kim1994gauge, Kim1995, polchinski1994low, lee1992gauge, oganesyan2001quantum, lee2008stability, fitzpatrick2014non, abanov2000spin, moon2013non, han2019emergent, nayak1994non, nayak1994renormalization, lee2009low, metlitski2010quantum1, metlitski2010quantum2, mross2010controlled, rech2006quantum, chubukov2004instability, dalidovich2013perturbative, lee2018recent, han2022non}, at quantum critical points associated with a change in volume
of the Fermi surface~\cite{coleman2001fermi, han2022non2, han2022microscopic, senthil2004weak, senthil2008critical, senthil2008theory}, or when fermions interact through random couplings~\cite{sachdev1993gapless, sachdev2010holographic, maldacena2016remarks, chowdhury2022sachdev, patel2023universal, Guo2021LargeI, Guo2022LargeII, guo2023largeIII}. As they are in a strong coupling regime, constructing a controlled theoretical framework to describe NFL has thus far remained an enticing but challenging objective despite intense interest.

Considering their fundamental role in strongly correlated electron systems, there is a crucial need for a more systematic understanding of NFL beyond perturbative treatments in more than one dimension. Substantial recent progress has been made in that regard. In particular, some universal constraints on the low-energy physics of metals --- that could apply for both FL and NFL --- have been derived using non-perturbative arguments coming from emergent symmetry groups and their corresponding anomalies~\cite{shi2022gifts, else2021non, else2021strange, lu2023definition, else2023collisionless, shi2023loop}.  Controlled expansions for critical Fermi surfaces based on random couplings in flavour space between the fermions and the bosonic order parameter have also been devised~\cite{Guo2021LargeI, Guo2022LargeII, guo2023largeIII, patel2023universal}. Most importantly for the purpose of our work, an effective field theory (EFT) for FL was put forward in Ref.~\cite{Son2022} based on a novel bosonization method for the Fermi surface. This work is the latest of the many historical attempts at bosonizing Fermi liquids in dimensions greater than one. Previous works have been based either on patch decomposition~\cite{haldane2005luttinger, Neto1994BosonizationPRB, Neto1994BosonizationPRL, houghton2000multidimensional, Houghton1995theory} or the dynamics of droplets in phase space akin to quantum Hall droplets on the lowest Landau level~\cite{Khveshchenko1994Bosonization, Khveshchenko1995Geometrical, iso1992fermions, karabali2004effective, polychronakos2005chiral, karabali2004edge}. Ref.~\cite{Son2022} showed that starting from any semiclassical phase space distribution $f^{\text{sc}}(\mathbf{p};t,\mathbf{x})$ that represents a sharp Fermi surface, all other possible Fermi liquid distributions can be obtained by a phase space volume-preserving transformation parametrized by a bosonic field $\phi(\theta;t,\mathbf{x})$ that depends on spacetime coordinates and the position on the Fermi surface $\theta$. An effective action that reproduces known results at quadratic order and systematically includes non-linear corrections was further proposed for these bosonic degrees of freedom. Constructing an EFT for FL that captures non-perturbative effects and allows one to calculate all desired observables at any given order is a significant step towards an analytical description of NFL. However, it remains \emph{a priori} unclear if the ideas put forward by Ref.~\cite{Son2022} can be adapted to describe the infrared behavior of NFL, whose physics is fundamentally different from FL. 

Inspired by this framework,  we propose a bosonized effective field theory for NFL. It is first shown that, in light of the quantum Boltzmann equation (QBE), one should use the frequency-dependent generalized distribution $f(\omega,\theta;t,\mathbf{x})$ rather than the semiclassical phase space distribution $f^{\text{sc}}(\mathbf{p};t,\mathbf{x})$ for the dynamics of NFL. Here, $\theta$ represents the angular coordinate along the FS in momentum space and $\omega$ the frequency. Indeed, it is highlighted that the usual distribution function over phase space does not satisfy a closed equation of motion in the absence of quasiparticles. In contrast, the generalized distribution $f(\omega,\theta;t,\mathbf{x})$ importantly captures the nontrivial frequency dependence inherent to any description of NFL and does follow a closed equation of motion even in the absence of sharp excitations. We then argue that physical fluctuations of these generalized distributions can be parametrized by a bosonic field and introduce an effective action. We demonstrate that this action yields the collisionless QBE as its equation of motion. Expanding the action to quadratic order shows that our effective theory reproduces known results for the case of a two-dimensional spherical Fermi surface. The action shows forward scattering cancellation, which can usually be attributed to a Ward identity~\cite{kim1994gauge, Kim1995}. By performing an angular momentum decomposition of the bosonic field, it is further proven that the small angular momentum modes display the usual Fermi liquid-like behavior. In contrast, the large angular momentum modes (or rough fluctuations) have a singular behavior characteristic of NFL --- a known result typically obtained through a detailed analysis of the QBE~\cite{Kim1995, Mandal2022}. This effective action thus seemingly captures highly nontrivial known results even if restricted to quadratic order. It should be emphasized that even though we focus in this article on the quadratic action for a spherical Fermi surface in two spatial dimensions, our proposed effective action applies to other nontrivial Fermi surfaces in any dimension and, most importantly, systematically captures non-linear corrections that would be intractable with other means. By offering a comprehensive framework to describe the physics of NFL analytically, our work could serve as a cornerstone towards a more complete understanding of strange metals and related problems. 

The rest of the paper is organized as follows: Section~\ref{sec:QBE_GDF} briefly reviews the derivation of the quantum Boltzmann equation and introduces the generalized distribution function. In Section~\ref{sec:effective_action}, we derive the effective action before expanding it to the quadratic order in terms of the bosonic field and computing response functions at first order in the self-energy expansion in Section~\ref{sec:quadatic}. We then comment on how to generalize the quadratic action using the full functional expansion of the self-energy in Section~\ref{sec:general}. In Section~\ref{sec:discussion}, we finally end by discussing our work's implications and possible future directions.

\section{Quantum Boltzmann equation and generalized distribution function}\label{sec:QBE_GDF}

Much of the current understanding of FL and NFL relies on using the quantum Boltzmann equation. We here introduce the usual derivation of these equations of motion before discussing how they can naturally emerge from an effective field theory in the following sections. 

The QBE describes the evolution of the fermionic distribution function over time and can be thought of as a coarse-grained version of the equation of motion for the non-equilibrium one-particle lesser Green's function
\begin{equation}
    G^{<}\left(\mathbf{x}_1, t_1; \mathbf{x}_2, t_2\right) = i\left\langle\psi^{\dagger}\left(\mathbf{x}_2, t_2\right) \psi\left(\mathbf{x}_1, t_1 \right)\right\rangle. 
\end{equation}
In the general non-equilibrium case, the Green's function does not have spatial or temporal translation symmetry. It depends on both the center of mass (i.e., $\mathbf{x}=(\mathbf{x}_1+\mathbf{x}_2)/2$ and $t=(t_1+t_2)/2$) and relative coordinates (i.e., $\mathbf{x}_{\mathrm{rel}}=\mathbf{x}_{2} - \mathbf{x}_1$ and $t_{\mathrm{rel}}=t_{2}-t_{1}$). 

Assuming one is interested in physics at length scales much larger than the inverse Fermi momentum  $1/p_F$ and on time scales much larger than the inverse Fermi energy $1/\varepsilon_F$, the equation of motion in the collisionless limit for the Fourier transform of the lesser Green's function with respect to the relative coordinates $G^{<}(\omega,\mathbf{p};t,\mathbf{x})$ is (see Appendix~\ref{app:Non-equilibirum_field_theory})~\cite{kita2010introduction, rammer1986quantum, rammer2011quantum}
\begin{equation}
\label{eq:Dyson_equation_general}
    [ \omega - \xi_{\mathbf{p}} -\Re\Sigma^{R}, G^{<}] - [\Sigma^{<}, \Re G^{R}] = 0,
\end{equation}
where we have introduced the free-particle spectrum $\xi_{\mathbf{p}}=\epsilon(\mathbf{p})-\mu$, the retarded Green's function $G^{R}$, its corresponding self-energy $\Sigma^{R}$, and the generalized Poisson bracket 
\begin{align} \label{eq:generalized_Poisson_bracket}
[B,C]=\frac{\partial B}{\partial\omega}\frac{\partial C}{\partial t}-\frac{\partial B}{\partial t}\frac{\partial C}{\partial\omega}-\frac{\partial B}{\partial p_{i}}\frac{\partial C}{\partial x_{i}}+\frac{\partial B}{\partial x_{i}}\frac{\partial C}{\partial p_{i}}
\end{align}
with the repeated indices summed over. This generalization contains additional derivatives for the relative frequency $\omega$ and center of mass time $t$ in contrast to the usual Poisson bracket. In equilibrium, the lesser Green's function takes the form~\cite{bruus2004many, negele2018quantum, mahan2000many}
\begin{equation}
    G^{<}( t,\mathbf{x}; \omega,\mathbf{p}) = i A( t,\mathbf{x};\omega,\mathbf{p}) f_{0}(\omega), 
\end{equation}
where $f_{0}(\omega)=\left( e^{\omega/T} + 1 \right)^{-1}$ is the Fermi-Dirac distribution and 
\begin{equation}
    A(\mathbf{p}, \omega)=\frac{-2 \operatorname{Im} \Sigma^R(\omega, \mathbf{p})}{\left(\omega-\xi_{\mathbf{p}}-\operatorname{Re} \Sigma^R(\omega, \mathbf{p})\right)^2+\left(\operatorname{Im} \Sigma^R(\omega, \mathbf{p})\right)^2}.
\end{equation}
is the spectral function. 

For Fermi liquids, quasiparticles remain well-defined even in the presence of strong interactions such that $\operatorname{Im} \Sigma^R\ll \omega$ at a sufficiently low energy scale. In this physically relevant low-energy limit, we can neglect the self-energy and approximate the quasiparticles to be infinitely long-lived~\cite{bruus2004many, negele2018quantum, mahan2000many}
\begin{align}
    A(\omega, \mathbf{p}) &\approx 2 \pi \delta\left(\omega-\xi_{\mathbf{p}}  \right). 
\end{align}
In the presence of quasiparticles, we can integrate out the frequency dependence of Eq.~\eqref{eq:Dyson_equation_general} to obtain a closed equation of motion  for the \emph{semiclassical distribution} function
\begin{equation}
    f^{\text{sc}}(\mathbf{p}; t, \mathbf{x}) = -i \int \frac{d \omega}{2 \pi} G^{<}(\omega,\mathbf{p}; t,\mathbf{x})
\end{equation}
which is the usual collisionless Boltzmann equation
\begin{equation} \label{eq:QBE_for_FL}
    \left[ \omega - \xi_{\mathbf{p}}, f^{\text{sc}} \right] = \partial_{t}f^{\text{sc}} +(\nabla_{\mathbf{p}}\epsilon(\mathbf{p}))\cdot(\nabla_{\mathbf{x}} f^{\text{sc}}) = 0. 
\end{equation}
A superscript ``sc'' emphasizes that this is analogous to the usual semiclassical phase space distribution.

In contrast, interactions are so important for NFL that the concept of quasiparticles breaks down since $\operatorname{Im} \Sigma^R$ scales as fast or faster than $\omega$ at low energy. The corresponding $\text{Re}\Sigma^{R}$ leads to a divergent effective mass of the quasiparticles. The spectral function is no longer a strongly peaked function of frequency, implying that, even near equilibrium, the frequency-independent distribution function $f^{\text{sc}}(\mathbf{p}; t, \mathbf{x})$ does not satisfy a closed equation of motion anymore.  Even in the absence of a sharp quasiparticle peak in $\omega$, the spectral function is still a well-peaked function of $\xi_{\mathbf{p}}$ at the Fermi surface for small frequency~\cite{Kadanoff1964,Kim1995}. In such cases, the momenta can be restricted to be on the Fermi surface in a procedure commonly referred to as the Prange-Kadanoff (PK) reduction~\cite{Kadanoff1964,Guo2022LargeII, tulipman2023criterion}.
An equation of motion can then be obtained by integrating Eq.~\eqref{eq:Dyson_equation_general} over $\xi_{\mathbf{p}}$ for the \emph{generalized distribution} function
\begin{equation}
    f( t, \mathbf{x}; \omega,\theta)=-i \int \frac{d \xi_\mathbf{p}}{2 \pi} G^{<}( t,\mathbf{x}; \omega,\mathbf{p}),
\end{equation}
where $\theta$ parametrizes the direction of $\mathbf{p}$. The equation of motion for the generalized distribution function is the collisionless QBE  
\begin{equation}\label{eq:QBE_for_NFL}
    \left[\omega-\xi_{\mathbf{p}}-\Sigma^{R}[f],f \right] = 0,
\end{equation}
where we used $\int d\xi_{\mathbf{p}} \Re
G^{R}=0$ (see Appendix~\ref{app:useful}) and consider the linearized dispersion \begin{equation} \label{eq:linearized_dispersion}
    \xi_{\mathbf{p}} = \epsilon(\mathbf{p})-\mu \approx v_{F}(\theta) p_{\parallel}. 
\end{equation}
Here, $p_\parallel$ is the momentum component parallel to $\nabla_{\mathbf{p}}\epsilon$ (i.e., perpendicular to the Fermi surface), and the Fermi velocity is $v_{F}(\theta) \mathbf{n}_{\theta}=\left.\nabla_{\mathbf{p}} \epsilon \right|_{\theta, \xi_p=0}$, where $\mathbf{n}_{\theta}$ is a unit vector perpendicular to the Fermi surface at a position parametrized by $\theta$. Notice that $\nabla_{\mathbf{p}}f\approx p_{F}^{-1}\mathbf{s}_{\theta}\partial_{\theta}f$, where $\mathbf{s}_{\theta}\cdot\mathbf{n}_{\theta}=0$, since $f$ only depends on the angle $\theta$ or the momentum transfer parallel to the Fermi surface. From this point onward, we will not mention that we are taking the real part to lighten the notation as in Eq.~\eqref{eq:QBE_for_NFL}. The generalized distribution can be thought of as describing the variation of the local chemical potential in momentum space.

\section{Effective action via generalized canonical transformation}\label{sec:effective_action}

The equations of motion described in Eqs.~\eqref{eq:QBE_for_FL} and \eqref{eq:QBE_for_NFL} offer important insight into the properties of FL and NFL. They allow for the calculation of certain response functions at leading order for small momentum. However, to gain a deeper and more systematic understanding, it would be preferable to have a description of the low-energy behavior of these phases of matter in terms of an effective field theory with the QBE as its saddle point solution. An effective action would provide a way to evaluate, in theory, all possible observables at any given order. For FL, such an effective action was recently proposed by Delacrétaz, Du, Mehta, and Son~\cite{Son2022} using the method of coadjoint orbits. Inspired by their work, we propose a bosonization approach suitable for NFL.

\subsection{Generalized canonical transformation}

The quantum Boltzmann equations~\eqref{eq:QBE_for_FL} and~\eqref{eq:QBE_for_NFL} govern the evolution of any semiclassical $f^{\text{sc}}$ and generalized distribution functions $f$, respectively. Both FL and NFL possess sharp Fermi surfaces at zero temperature, which persist even when the distribution undergoes time evolution under the collisionless QBE. Consequently, it suffices to consider these physical distributions with sharp Fermi surfaces to capture the fluctuations of FL and NFL. In the case of Fermi liquids, this corresponds to distributions that exhibit a jump at the Fermi momentum from the occupied to the unoccupied side. For NFL, generalized distributions with Fermi surfaces are associated with Green's functions that manifest a singularity or kink at the Fermi momentum in their momentum occupation $n(\mathbf{p})$. By starting from any representative $f_0$ of this class of distribution functions, other physical distributions $f$ with a Fermi surface can be parametrized by a transformation $U$. 

One of the main idea behind the coadjoint orbit formalism of Ref.~\cite{Son2022} is then to identify the space of physical states of interest (i.e., semiclassical distributions with sharp Fermi surfaces) as the space of distributions that can be obtained by canonical transformations $U\in\mathcal{G}_{\text{sc}}$ on phase space starting from a valid distribution such as $f_0^{\text{sc}}(\mathbf{x},\mathbf{p})=\Theta(\abs{\mathbf{p}}- p_F)$. This space is formally defined as the orbit of $f_0$ under the group of canonical transformation $\mathcal{G}_{\text{sc}}$ over phase space (see Appendix~\ref{app:math_details} for mathematical details). Within this space, one can further identify a one-to-one correspondence between physical distributions $f^{\text{sc}}(\mathbf{x},\mathbf{p})$ and fields $\phi$, which are generators of canonical transformations (i.e., elements of the Lie algebra $\mathfrak{g}_{\text{sc}}$ of canonical transformations). Given such a correspondence, a field theory for $\phi$ directly describes fluctuations of the Fermi surface in phase space.

The above ideas can be adapted to generalized distribution functions with additional dependence on time $t$ and frequency $\omega$. In this case, the simplest initial distribution to describe a sharp Fermi surface is $\tilde{g}_0(t,\mathbf{x};\omega,\mathbf{p})=2\pi\delta(\omega-\xi_\mathbf{p})\Theta(-\omega)$, where $\Theta(-\omega)$ is the Fermi-Dirac distribution at zero temperature. Other relevant generalized distributions at zero-temperature can be obtained by acting with a generalized transformation $U=\exp\left( -\phi\right)\in\mathcal{G}$ as
\begin{align} \label{eq:canonical_transformatioon_quasiclassical_distribution}
    \tilde{g}(t,\mathbf{x}; \omega,\mathbf{p})=&\tilde{g}_{0}-[\phi,\tilde{g}_{0}]+\frac{1}{2!}[\phi,[\phi,\tilde{g}_{0}]]+\ldots \notag\\
    =& U \tilde{g}_{0} U^{-1},
\end{align}
where the generalized Poisson bracket introduced in Eq.~\eqref{eq:generalized_Poisson_bracket} has to be used since the distribution $\tilde{g}$ depends on frequency, time, position and momentum. We note that $U$ is not the exponential of a function but rather the exponential map of a Lie algebra element $\phi\in\mathfrak{g}$, which is isomorphic to a smooth function  $\phi\cong\phi(t, \mathbf{x};\omega,\mathbf{p})$. Eq.~\eqref{eq:canonical_transformatioon_quasiclassical_distribution} then naturally follows from the Baker–Campbell–Hausdorff formula~\cite{hall2013lie, miller1973symmetry}. The Lie algebra $\mathfrak{g}$ should be thought of as the space of physical observables. The deviation of the generalized distribution function from the equilibrium $\delta \tilde{g}=\tilde{g}-\tilde{g}_{0}$ can be written in terms of $\phi$ as $\delta \tilde{g}=\sum_{n=1}\Delta_{\phi}^{(n)}\tilde{g}$ where
\begin{align}\label{eq:expansion_qc_distribution_phi}
\Delta^{(n)}_{\phi}\tilde{g}={}&\frac{(-1)^{n}}{n!}[\underbrace{\phi,[\phi,\ldots,[\phi}_{n},\tilde{g}_{0}]\ldots]].
\end{align}

At this stage, one does not have a unique correspondence between distributions $\tilde{g}$ that can be obtained from $\tilde{g}_0$ through generalized canonical transformations and Lie algebra elements $\phi$. Indeed, there is a huge redundancy in the description since any two transformations $U=e^{-\phi}$ and $UV$ will yield the same state $\tilde{g}=U\tilde{g}_0U^{-1}$ for any $V=e^{\alpha}$ that leaves the initial distribution $\tilde{g}_{0}$ invariant (i.e., $V\tilde{g}_0 V^{-1}=\tilde{g}_0$). $V$ is said to be a member of the stabilizer or little group of $\tilde{g}_{0}$. To get a unique identification, we need to pick a single representative from the equivalence class defined by the relation
\begin{align}
    e^{-\phi} & \sim e^{-\phi} e^{\alpha}  \Longrightarrow \phi \sim \phi-\alpha+\frac{1}{2}\{\phi, \alpha\}+\ldots. \label{eq:gauge_redundancy_distribution}
\end{align}
This can be achieved through a ``gauge fixing'' procedure. A suitable choice is to remove the dependence of $\phi$ on the momentum amplitude 
\begin{align}\label{eq:gauge_fixing_condition}
    \phi = \left.\phi(t,\mathbf{x};\omega,\theta)\right|_{\xi_{\mathbf{p}}=\omega}.
\end{align}
This can be seen by noting that any distribution can be brought to this form at linear order using a gauge transformation of the form \eqref{eq:gauge_redundancy_distribution} with 
\begin{align}
    \alpha(t,\mathbf{x};\omega,\mathbf{p}) = \phi(t,\mathbf{x};\omega,\mathbf{p}) - \left.\phi(t,\mathbf{x};\omega,\theta)\right|_{\xi_p=\omega},
\end{align}
which is a valid transformation since $\comm{\alpha}{\tilde{g}_0}=0$ (i.e., $e^{\alpha} \tilde{g}_0 e^{-\alpha}=\tilde{g}_0$).

\subsection{Self-energy expansion and Prange-Kadanoff reduction} \label{subsec:SE_and_PK_reduction}

A central object in the equation of motion for NFL is the self-energy. $\Sigma^{R}$ is a functional of the generalized distribution $\tilde{g}$ and can be expanded as
\begin{align}
\tilde{\Sigma}^{R}[\tilde{g}] =&\int_{\bar{\mathbf{1}}} \tilde{D}_1\left(t,\mathbf{x}, \omega, \mathbf{p}; \bar{\mathbf{1}} \right) \tilde{g}\left(\bar{\mathbf{1}}\right) \nonumber \\
&+  \int_{\bar{\mathbf{1}}, \bar{\mathbf{2}}} \tilde{D}_2\left(t,\mathbf{x}, \omega, \mathbf{p}; \bar{\mathbf{1}}; \bar{\mathbf{2}}\right) \tilde{g}\left(\bar{\mathbf{1}}\right) \tilde{g}\left(\bar{\mathbf{2}}\right)+\ldots \nonumber \\
=& \sum_{n=1}^{\infty} \tilde{\Sigma}^{R(n)}[\tilde{g}], \label{eq:self_expand}
\end{align}
where we have used the notation $\bar{\mathbf{1}}\equiv\left(t_1, \mathbf{x}_1,\omega_1,\mathbf{p}_1\right)$ and $\int_{\bar{\mathbf{1}}}=\int \frac{dt_{1} d^{d}x_{1} d\omega_{1} d^{d}p_{1}}{(2\pi)^{d+1}}$. The coefficients
\begin{align}
    \tilde{D}_n\left(t,\mathbf{x}, \omega, \mathbf{p}; \bar{\mathbf{1}}; \ldots ; \bar{\mathbf{n}} \right)=\frac{1}{n!} \frac{\delta^n\tilde{\Sigma}^R}{\delta \tilde{g}\left(\bar{\mathbf{1}}\right) \ldots \delta \tilde{g}\left(\bar{\mathbf{n}}\right)}.
\end{align}
are the generalized Landau-interaction parameters. In the following, we assume that these generalized coefficients do not depend on position or time, are invariant under the change of variables 
\begin{equation} \label{eq:Change_of_variables_Landau_parameters}
    \tilde{D}_{n}(\omega,\mathbf{p}, \ldots, \omega_j,\mathbf{p}_j,\ldots) = \tilde{D}_{n}(\omega_j,\mathbf{p}_j,\ldots,\omega,\mathbf{p}, \ldots),
\end{equation} 
and only depend on the frequency and momentum differences $\tilde{D}_{n}(\omega,\mathbf{p},\omega_{1},\mathbf{p}_{1},\ldots)=\tilde{D}_{n}(\omega_{1}-\omega,\mathbf{p}_{1}-\mathbf{p};\ldots;\omega_{n}-\omega,\mathbf{p}_{n}-\mathbf{p})$. With these assumptions, the $n^{\text{th}}$-order term in the self-energy expansion $\tilde{\Sigma}^{R(n)}$ is
\begin{widetext}
\begin{align}\label{eq:nth_order_self_energy}
\tilde{\Sigma}&^{R(n)}[\tilde{g}]=
\int_{\omega_{1}\mathbf{p}_{1}}\ldots\int_{\omega_{n}\mathbf{p}_{n}}\tilde{D}_{n}(\omega_{1}-\omega,\mathbf{p}_{1}-\mathbf{p};\ldots;\omega_{n}-\omega,\mathbf{p}_{n}-\mathbf{p})
\prod_{i=1}^{n}\tilde{g}(t,\mathbf{x};\omega_{i},\mathbf{p}_{i}).
\end{align}
Given the required assumptions are met, one can further use the PK reduction by fixing the momenta in the generalized coefficients to be on the Fermi surface
\begin{align}\label{eq:nth_order_landau_coefficients_after_PK_reduction}
    &\tilde{D}_{n}(\omega_{1}-\omega,\mathbf{p}_{1}-\mathbf{p};\ldots;\omega_{n}-\omega,\mathbf{p}_{n}-\mathbf{p})\approx \left.D_{n}(\omega_{1}-\omega,\theta_{1}-\theta;\ldots;\omega_{n}-\omega,\theta_{n}-\theta)\right|_{\xi_\mathbf{p}=0,\ldots,\xi_{\mathbf{p}_n}=0}.
\end{align}
In this case, the $n^{\text{th}}$-order term in the self-energy expansion becomes
\begin{align}\label{eq:nth_order_self_energy_after_PK_reduction}
    \Sigma&^{R(n)}[g]=\int_{\omega_{1}\theta_{1}}\ldots\int_{\omega_{n}\theta_{n}}D_{n}(\omega_{1}-\omega,\theta_{1}-\theta;\ldots;\omega_{n}-\omega,\theta_{n}-\theta) \prod_{i=1}^{n} g(t,\mathbf{x};\omega_{i},\theta_{i}),
\end{align}
\end{widetext}
where we have introduced 
\begin{align}
    g(t,\mathbf{x};\omega,\theta) = \int \frac{d\xi}{2\pi} N(\xi,\theta) \tilde{g}(t,\mathbf{x};\omega,\xi,\theta).
\end{align}
The angular-resolved density of states is defined such that $\int_{\mathbf{p}}=\int d^d p=\int d^{d-1}\theta d\xi N(\xi,\theta)$. Assuming the density of states only has weak momentum dependence that does not significantly affect the low-energy dynamics of the system, we will fix it to its values at the Fermi energy in the rest of the article $N(\xi,\theta)\approx N(0,\theta)$. The associated initial distribution is simply the zero-temperature Fermi-Dirac distribution weighted by the density of states $g_0=\int \frac{d\xi}{2\pi} N(0,\theta) \tilde{g}_0 = N(0,\theta) \Theta(-\omega)$. It is important to note that, assuming an isotropic density of states $N(0,\theta)=N(0)$, we have 
\begin{align}\label{eq:distribution_post_energy_integration}
    g = \int \frac{d\xi}{2\pi} N(0) U \tilde{g}_0 U^{-1} = U g_0 U^{-1}
\end{align}
as a consequence of the gauge fixing condition~\eqref{eq:gauge_fixing_condition}. It follows that the deviation from the equilibrium distribution $\delta g=\sum_{n=1}\Delta_{\phi}^{(n)}g$ is identical to Eq.~\eqref{eq:expansion_qc_distribution_phi}
\begin{align}\label{eq:expansion_qc_distribution_phi_post_PK}
\Delta^{(n)}_{\phi}g={}&\frac{(-1)^{n}}{n!}[\underbrace{\phi,[\phi,\ldots,[\phi}_{n},g_{0}]\ldots]].
\end{align}
To make the notation as explicit as possible, we remove the tilde over the self-energy, its coefficients, and the generalized distribution after the PK reduction procedure. 

Given the above functional dependence on the distribution function, the self-energy displacement from its equilibrium configuration $\tilde{\Sigma}^{R}_0\equiv\tilde{\Sigma}^{R}[\tilde{g}_0]$ can also be expanded in powers of $\phi$ as
\begin{equation}\label{eq:expansion_self-energy_distribution_phi}
    \delta\tilde{\Sigma}^{R}[\tilde{g}]= \tilde{\Sigma}^{R}[\tilde{g}]- \tilde{\Sigma}^{R}_0 =\sum_{n=1}^{\infty}\sum_{m=0}^{\infty}\Delta^{(m)}_{\phi}\tilde{\Sigma}^{R(n)},
\end{equation}
where the generalized distribution is displaced from $\tilde{g}_0$ as in Eq.~\eqref{eq:expansion_qc_distribution_phi}. $\Delta_{\phi}^{(m)}\tilde{\Sigma}^{R(n)}$ is a shorthand notation for the sum of all terms that come from the $n^{\text{th}}$ order in the functional expansion of the self-energy $\tilde{\Sigma}^{R(n)}$ (see Eqs.~\eqref{eq:self_expand} and~\eqref{eq:nth_order_self_energy}) and include $m$ functions $\phi$. This notation can also be introduced for the self-energy after the PK reduction procedure by replacing the self-energy $\tilde{\Sigma}\to\Sigma$ and the generalized distribution $\tilde{g}\to g$, such that 
\begin{equation}\label{eq:expansion_self-energy_distribution_phi_post_pk}
    \delta\Sigma^{R}[g]= \Sigma^{R}[g]- \Sigma^{R}_0 =\sum_{n=1}^{\infty}\sum_{m=0}^{\infty}\Delta^{(m)}_{\phi}\Sigma^{R(n)},
\end{equation}
where $\Sigma^{R}_0\equiv\Sigma^{R}[g_0]$.

\subsection{Effective action for (Non-)Fermi liquids}

\subsubsection{General effective action}

We propose the following effective action to describe the low-energy physics of metals with and without quasiparticles
\begin{align}
\tilde{S}[\phi]={}& \tilde{S}_{\text{WZW}} + \tilde{S}_H \notag \\
={}& \braket{\tilde{g}_0,U\partial_t U^{-1}} - \braket{\tilde{g},\xi_{\mathbf{p}}+\sum_{n=1}\tfrac{1}{n+1}\tilde{\Sigma}^{R(n)}[\tilde{g}]}\notag\\
={}& \braket{\tilde{g},\omega-\xi_{\mathbf{p}}-\sum_{n=1}\tfrac{1}{n+1}\tilde{\Sigma}^{R(n)}[\tilde{g}]},\label{eq:action_before_PK}
\end{align}
where $\braket{B,C}$ is the inner product defined by
\begin{align}
    \braket{B,C}= \int\frac{dt d^{d}x d\omega d^{d}p}{(2\pi)^{d+1}} BC
\end{align}
and  $\xi_{\mathbf{p}}$ is the linearized dispersion at the Fermi surface as in Eq.~\eqref{eq:linearized_dispersion}. Eq.~\eqref{eq:action_before_PK} is an effective field theory for the bosonic degrees of freedom $\phi$ that parametrizes fluctuations of the generalized distribution.

To verify that the above effective action properly captures the dynamics of non-Fermi liquids, let us obtain the equation of motion $\delta S/\delta \phi=0$. To do so we vary the field $\phi$ by an infinitesimal amount $\alpha$ (i.e., $U\rightarrow e^{-\alpha}U$). The variations of the action at linear order in $\alpha$ are (see Appendix~\ref{app:general_action} for details)
\begin{subequations}
\begin{align}
\delta[U^{-1}BU]&=U^{-1}[\alpha,B]U,\label{eq:varA}\\
\delta[U^{-1}\tilde{\Sigma}^{R(n)}[\tilde{g}]U]&=U^{-1}[\alpha,\tilde{\Sigma}^{R(n)}]U \notag\\
& - n\int_{\omega_{1} \mathbf{p}_{1}}\ldots\int_{\omega_{n} \mathbf{p}_{n}}  \tilde{D}_{n}[\alpha_{n},\tilde{g}_{n}]_{n}\prod_{i=1}^{n-1}\tilde{g}_{i} ,\label{eq:varS}
\end{align}
\end{subequations}
where $B$ stands for any parameters in the action without implicit dependence on the generalized distribution function, $\alpha_{i}\equiv \alpha(t,\mathbf{x};\omega_i,\mathbf{p}_i)$ and $\tilde{g}_{i}\equiv \tilde{g}(t,\mathbf{x};\omega_i,\mathbf{p}_i)$ depend on $\omega_{i}$ and $\mathbf{p}_{i}$, and $[A,B]_{n}$ indicates that the derivatives in the generalized Poisson bracket are evaluated with respect to $\omega_{n}$ and $\mathbf{p}_{n}$. By using Eqs.~\eqref{eq:varA} and \eqref{eq:varS}, the first-order variation of the action $\delta \tilde{S}$ is given by
\begin{align}
\delta\tilde{S}
={}&\braket{[\omega-\xi_{\mathbf{p}}-\tilde{\Sigma}^{R}[\tilde{g}],\tilde{g}],\alpha},
\end{align}
where the identity $\braket{A,[B,C]}=\braket{[A,B],C}$ was used. 
Considering $\alpha$ does not depend on $\xi_{\mathbf{p}}$ from the gauge fixing condition~\eqref{eq:gauge_fixing_condition}, the equation of motion is
\begin{align}
\int \frac{d\xi}{2\pi} N(0,\theta) [\omega-\xi_{\mathbf{p}}-\tilde{\Sigma}^{R}[\tilde{g}],\tilde{g}]=0,\label{eq:QBE_before_PK_reduction}
\end{align}
which is the energy integral of the first term in the equation of motion for the lesser Green's function~\eqref{eq:Dyson_equation_general}. Applying the PK reduction by fixing the momentum dependence of the self-energy on the Fermi surface as in Eq.~\eqref{eq:nth_order_self_energy_after_PK_reduction} and assuming $N(0,\theta)=N(0)$ yields
\begin{align}
[\omega-\xi_{\mathbf{p}}-\Sigma^{R}[g], g]=0.\label{eq:QBE_after_PK_reduction}
\end{align}
This is none other than the collisionless QBE introduced in Eq.~\eqref{eq:QBE_for_NFL}.

In the present formalism, once the effective action is specified, the physically correct procedure is to perform the path integral over all configurations of the bosonic field $\phi$, with an appropriate regularization. This is because $\phi$ encodes quantum fluctuations of the fermionic system and is not intended to be fixed to a single classical configuration. Treating $\phi$ as a classical field would correspond to taking the saddle-point approximation of the effective action, which recovers the collisionless quantum Boltzmann equation but neglects quantum fluctuations.
\subsubsection{Effective action post-Prange-Kadanoff reduction}

Rather than applying the PK reduction at the very end, one can directly apply it at the level of the action~\eqref{eq:action_before_PK}. In this case, one has
\begin{align}
    S[\phi]={}&  \langle \tilde{g}, \omega-\xi_{\mathbf{p}}-\sum_{n=1}\tfrac{1}{n+1}\Sigma^{R(n)}[g]\rangle \label{eq:action} 
\end{align}
Following an analogous reasoning to the previous derivation and using  Eqs.~\eqref{eq:distribution_post_energy_integration} and~\eqref{eq:expansion_qc_distribution_phi_post_PK}, the equation of motion can be shown once again to be the collisionless QBE~\eqref{eq:QBE_after_PK_reduction}. In the rest of the article, we will focus on the effective action post-PK reduction~\eqref{eq:action}. However, we stress that in situations where one suspects the momentum dependence of the self-energy away from the Fermi surface to be important, the effective action~\eqref{eq:action_before_PK} and its associated equation of motion~\eqref{eq:QBE_before_PK_reduction} are still applicable. 

\subsubsection{Luttinger-Ward functional}

At first glance, the factor on the right of the inner product in the action~\eqref{eq:action} might look similar to the inverse Green's function. However, there is an extra factor of $(n+1)^{-1}$ in front of the $n^{\text{th}}$ order in $\tilde{g}$ of the self-energy expansion. The factor is required to reproduce the quantum Boltzmann equation from the action~\eqref{eq:action} since the functional dependence of the self-energy on the distribution function leads to extra contributions (see  Eq.~\eqref{eq:varS}). Eq.~\eqref{eq:action} can be rewritten as
\begin{align}\label{eq:action_luttinger-Ward}
    S={}& \braket{\tilde{g}_0,\omega-\xi_{\mathbf{p}}} - \int_{tx}\Phi[g],
\end{align}
where we have introduced the functional
\begin{align}\label{eq:definition_generalized_luttinger-Ward}
\Phi[g]={}\sum_{n=1}^{\infty}\frac{1}{n+1}\int_{\omega\theta}\Sigma^{R(n)}[g] g.  
\end{align}
Interestingly, the functional derivative of $\Phi[f]$ with respect to the distribution satisfies $\delta\Phi/\delta g=\Sigma^{R}$, which is completely analogous to the conventional Luttinger-Ward functional~\cite{luttinger1960ground, baym1961conservation, kotliar2006electronic}. This analogy can even be pushed further by noting that the Feynman diagrams produced by this term are in correspondence with some two-particle irreducible bubble diagrams (i.e., ``skeleton'' diagrams~\cite{potthoff2004non, kozik2015nonexistence}) of the initial fermion theory at equilibrium (see Appendix~\ref{app:LW} for details). Considering the analogy, we shall refer to $\Phi$ as the generalized Luttinger-Ward functional of our effective field theory. 

In sum, the proposed action~\eqref{eq:action} produces the collisionless quantum Boltzmann equation.  
It does not assume the existence of well-defined quasiparticles and captures the non-trivial frequency dependence of NFL through the self-energy and its use of the generalized distribution function. It should be underscored that this action contains terms at all orders in $\phi$.

\section{Quadratic theory at one loop order}\label{sec:quadatic}

The action proposed in the last section describes a non-linear effective theory for the $\phi$ field that parametrizes fluctuations of the generalized distribution around its equilibrium position. In this section, we discuss how to expand the effective action in terms of $\phi$ and use it to compute observables. In particular, we will study the quadratic action in $\phi$ assuming a spherical Fermi surface in two dimensions to compare with existing results. This implies that the angular-resolved density of states and Fermi velocity are now isotropic (i.e., $N(0,\theta)=N(0)$,  $v_{F}(\theta)=v_{F}$, and $\mathbf{n}_{\theta}=\mathbf{p}_{F}/p_{F}$). In addition, we will consider the first-order expansion in $g$ for the self-energy (i.e., one-loop approximation in the corresponding diagrammatic perturbation theory)
\begin{align}\label{eq:First_order_sefl_energy}
    \Sigma^{R}[g]={}&\int_{\omega'\theta'} D_{1}(\omega' -\omega;\theta'-\theta) g(t,\mathbf{x},\omega',\theta') = \Sigma^{R(1)}[g].
\end{align}
On top of assuming the PK reduction is valid, we further suppose that the first order generalized coefficient above is invariant under the change of variables $D_{1}(\omega'-\omega,\theta'-\theta)=D_{1}(\omega-\omega',\theta'-\theta)=D_{1}(\omega'-\omega,\theta-\theta')\equiv D_1$ (as stated in Sec.~\ref{subsec:SE_and_PK_reduction}). Such assumptions are accurate in many physical situations of interest like a Fermi surface in two dimensions interacting with a transverse gauge boson~\cite{kim1994gauge,Kim1995} or a critical Ising-nematic order parameter~\cite{Mandal2022}. In the random phase approximation (RPA) to these cases, the boson propagator $D_{1}$ is approximately given by $D_{1}(\omega'-\omega,\mathbf{p}'-\mathbf{p})\approx D_{1}(\omega'-\omega,p_{F}|\theta'-\theta|)$ since the main contribution comes from near the Fermi surface. We emphasize that the action can be used for more general Fermi surfaces in any dimension and systematically captures higher-order contributions in $\phi$ for the effective action and in $f$ for the self-energy. The general case, which includes the higher-order contributions in $g$ for the self-energy, will be discussed in the next section. The current section should be seen as the lowest-order approximation to the action~\eqref{eq:action}, and the first step in obtaining an improved bosonized action.

\subsection{Derivation of the Gaussian action}\label{subsec:quadatic_gaussian}

Using the generalized canonical transformations for the generalized distribution~\eqref{eq:canonical_transformatioon_quasiclassical_distribution} as well as the definitions given in Eqs.~\eqref{eq:expansion_qc_distribution_phi} and~\eqref{eq:expansion_self-energy_distribution_phi_post_pk} for the distribution and self-energy displacement, the action within the one-loop approximation can be expanded as 
\begin{widetext}
\begin{align}
S={}& \langle \tilde{g}, \omega-\xi_{\mathbf{p}}-\tfrac{1}{2}\Sigma^{R}[g]\rangle\notag\\
={}& \langle \tilde{g}_{0}+\Delta_{\phi}^{(1)}\tilde{g} + \Delta_{\phi}^{(2)}\tilde{g} + \ldots,\omega-\xi_{\mathbf{p}}-\tfrac{1}{2}(\Sigma^{R}[g_0]+\Delta_{\phi}^{(1)}\Sigma^{R}+\Delta_{\phi}^{(2)}\Sigma^{R}+\ldots)\rangle \notag\\
={}& \langle \tilde{g}_{0}, \omega-\xi_{\mathbf{p}}-\tfrac{1}{2}\Sigma^{R}_{0}\rangle + \langle\Delta_{\phi}^{(1)}\tilde{g},\omega-\xi_{\mathbf{p}}-\Sigma^{R}_{0}\rangle + \langle \Delta_{\phi}^{(2)}\tilde{g}, \omega-\xi_{\mathbf{p}}-\Sigma^{R}_{0}\rangle
-\langle\Delta_{\phi}^{(1)}\tilde{g},\tfrac{1}{2}\Delta_{\phi}^{(1)}\Sigma^{R}\rangle +\ldots,\label{eq:effective_expand}
\end{align}
\end{widetext}
where $\Sigma_{0}^{R}=\Sigma^{R(1)}[g_{0}]$, $\Delta_{\phi}^{(m)}\Sigma^{R}=\Delta_{\phi}^{(m)}\Sigma^{R(1)}$, and we used $ \braket{A,[B,C]}= \braket{[A,B],C}$ and $ \braket{\Delta_{\phi}^{(m)} \tilde{g} ,\Delta_{\phi}^{(n)}\Sigma^{R}}
= \braket{\Delta_{\phi}^{(n)} \tilde{g} ,\Delta_{\phi}^{(m)}\Sigma^{R}}$  (see Appendix~\ref{app:useful} for details). The first and second terms of the last equality correspond to the $\phi^{0}$ and $\phi^{1}$ orders, respectively. The action at the $\phi^{0}$ order is a constant, and the $\phi^{1}$ order contribution gives a vanishing boundary term. The third term is the quadratic action $S^{(2)}$, whereas the ellipsis stands for higher-order corrections that we discuss in more detail in Appendix~\ref{app:higher}. 

The Gaussian action for $\phi$, $S^{(2)}$, can be split into two: the non-interacting $S^{(2)_{0}}$ and interacting $S^{(2)_{\text{int}}}$ parts
\begin{subequations}
\begin{align}
S^{(2)}={}&S^{(2)_{0}}+S^{(2)_{\text{int}}},\\
S^{(2)_{0}}={}&-\frac{1}{2}\int_{tx\omega\theta}(\partial_{\omega}g_{0})\dot{\phi}[\dot{\phi}+v_{F}\mathbf{n}_{\theta}\cdot(\nabla_{\mathbf{x}}\phi)],\label{eq:cont_quad_bare}\\
S^{(2)_{\text{int}}}={}&\frac{1}{2}\int_{tx\omega\theta}(\partial_{\omega}g_{0})\dot{\phi}[(\partial_{\omega}\Sigma^{R}_{0})\dot{\phi}-\Delta^{(1)}_{\phi}\Sigma^{R}].\label{eq:cont_quad_int}
\end{align}
\end{subequations}

We can check the equation of motion for the quadratic Gaussian action. The linear order variation for $\phi\rightarrow\phi+\alpha$ is
\begin{align}
    \delta S^{(2)}={}&\braket{[\omega-\xi_{\mathbf{p}}-\Sigma^{R}_{0},\Delta_{\phi}^{(1)}\tilde{g}]-[\Delta_{\phi}^{(1)}\Sigma^{R},\tilde{g}_{0}],\alpha},
\end{align}
which leads to the equation of motion 
\begin{align}
    [\omega-\xi-\Sigma^{R}_{0},\Delta_{\phi}^{(1)}g]-[\Delta_{\phi}^{(1)}\Sigma^{R},g_{0}]=0.
\end{align}
This is simply the QBE for NFL expanded at the $\phi^{1}$ order.

In terms of the Fourier transformed field $\hat{\phi}(\Omega,\mathbf{q};\omega,\theta)=\int_{tx}e^{-i(\Omega t-\mathbf{q}\cdot\mathbf{x})}\phi(t,x;\omega,\theta)$, the two parts of the Gaussian actions are
\begin{widetext}
\begin{subequations}\label{eq:disc_quad}
\begin{align}
    S^{(2)_{0}}={}&-\frac{N(0)}{2}\int_{\Omega q\omega\theta}F_{0}(\omega,\Omega)(\Omega-v_{F}(\mathbf{n}_{\theta}\cdot\mathbf{q}))\hat{\phi}(\Omega,\mathbf{q};\omega,\theta)\hat{\phi}(-\Omega,-\mathbf{q};\omega,\theta),\label{eq:disc_quad_bare}\\
   S^{(2)_\text{int}}
    ={}&\frac{N(0)}{2}\int_{\Omega q\omega\theta}F_{0}(\omega,\Omega)(\Sigma^{R}_{0}(\omega+\Omega)-\Sigma^{R}_{0}(\omega))\hat{\phi}(\Omega,\mathbf{q};\omega,\theta)\hat{\phi}(-\Omega,-\mathbf{q};\omega,\theta)\notag\\
    &-\frac{N(0)^2}{2}\int_{\Omega q\omega\theta}\int_{\omega'\theta'}F_{0}(\omega,\Omega)F_{0}(\omega',\Omega)D_{1}(\omega'-\omega,\theta'-\theta)\hat{\phi}(\Omega,\mathbf{q};\omega,\theta)\hat{\phi}(-\Omega,-\mathbf{q};\omega',\theta')\label{eq:disc_quad_int}
\end{align}
\end{subequations}
where we used the fact that $g_0(\omega)=N(0)f_0(\omega)$, discretized the frequency derivative ($(\partial_{\omega}f_{0})\Omega\approx f_{0}(\omega+\Omega)-f_{0}(\omega)$ for small $\Omega$) for regularization (see Appendix~\ref{app:quadratic} for details) and used a shorthand notation $F_{0}(\omega,\Omega)\equiv f_{0}(\omega+\Omega)-f_{0}(\omega)$. Using Eq.~\eqref{eq:First_order_sefl_energy} for the equilibrium self-energy, the interacting part $S^{(2)_{\text{int}}}$ can be rewritten as
\begin{align}
S^{(2)_{\text{int}}}={}&\frac{N(0)^2}{2}\int_{\Omega q\omega\theta}\int_{\omega'\theta'}F_{0}(\omega,\Omega)F_{0}(\omega',\Omega)D_{1}(\omega'-\omega,\theta'-\theta)\hat{\phi}(\Omega,\mathbf{q};\omega,\theta)
[\hat{\phi}(-\Omega,-\mathbf{q};\omega,\theta)-\hat{\phi}(-\Omega,-\mathbf{q};\omega',\theta')].
\end{align}
\end{widetext}
By inspection, one can notice that the interacting part of the quadratic action vanishes for forward scattering (i.e., near $\omega=\omega'$ and $\theta=\theta'$). In such a regime, the system thus exhibits Fermi liquid behavior. In the study of a Fermi surface coupled to a transverse gauge boson, it has already been pointed out that forward scattering cancellation arises from a Ward identity~\cite{kim1994gauge} and can be observed in the QBE~\cite{Kim1995}. The simplicity by which this non-trivial result can be obtained with our formalism graciously demonstrates its usefulness. In the following subsections, we shall study this quadratic action in detail to see how its predictions compare with existing results from the literature. 

\subsection{Density operator and algebra of densities}

In the next subsections, we will be interested in the charge response predicted by the quadratic action~\eqref{eq:disc_quad}. Before going through with such calculations, one should first wonder how to properly define the density operator in our formalism. Being an observable, the density operator should be an element of the Lie algebra $\mathfrak{g}$, which can be represented by a smooth function over the generalized phase space. In particular, the usual equal-time angular density is
\begin{align}
    \rho(t_1,\mathbf{x}_1,\theta_1) = \delta(t-t_1) \delta^{d}(\mathbf{x}-\mathbf{x}_1) \delta^{d-1}(\theta - \theta_1).
\end{align}
The density operator can be expanded in terms of the field $\phi$ by representing it as an operator acting on the Hilbert space of the EFT (see Appendix~\ref{app:density_and_algebra})
\begin{align}
    \rho[\phi](t,\mathbf{x},\theta) &= \left\langle \tilde{g}(t_1,\mathbf{x}_1;\omega_1,\mathbf{p}_1) , \rho(t_1, \mathbf{x}_1,, \theta_1) \right\rangle \nonumber \\
    &= \int \frac{d\omega dp p^{d-1}}{(2\pi)^{d+1}} U\tilde{g}_0(t,\mathbf{x};\omega,\mathbf{p})U^{-1}\nonumber \\
    &= \int \frac{d\omega dp  p^{d-1}}{(2\pi)^{d+1}} \left( \tilde{g}_0 -\comm{\phi}{\tilde{g}_0} + \ldots \right). \label{eq:def_density_operator}
\end{align}
With these density operators, we can define their commutator. The commutator between two densities, when considered as operators acting on the Hilbert space of EFT, is defined via the Lie bracket in $\mathfrak{g}$ through the relation $\comm{.}{.}_{\text{Q}}\to i\comm{.}{.}$, where the symbol on the left (right) is the commutator (Lie bracket). The density commutator expanded in powers of $\phi$ is then (see Appendix~\ref{app:density_and_algebra})
\begin{align}
    &\comm{\rho[\phi](t,\mathbf{x},\theta)}{\rho[\phi](t',\mathbf{x}',\theta')}_{\text{Q}} \nonumber \\
    &= \left\langle \tilde{g}, i \comm{\rho(t,\mathbf{x},\theta)}{\rho(t',\mathbf{x}',\theta')} \right\rangle \nonumber \\
    &= - i \frac{ p_F^{d-1}}{(2\pi)^{d}}  \mathbf{n}_{\theta}\cdot \nabla_{\mathbf{x}}\delta^{d}\left(\mathbf{x}-\mathbf{x}'\right) \delta^{d-1}\left(\theta-\theta'\right) \delta\left(t-t' \right)   + \ldots,\label{eq:density_algebra}
\end{align}
where the ellipsis stand for terms $\mathcal{O}(\phi)$. The commutation relation above recovers the algebra density used as a starting point in many multidimensional bosonization approaches~\cite{Neto1994BosonizationPRB, Neto1994BosonizationPRL, houghton1993Bosonization, houghton2000multidimensional} and, at leading order, is the same as in the Fermi surface bosonization framework of Ref.~\cite{Son2022}.

\subsection{Two-point density correlation function in non-interacting limit} \label{subsec:charge_reponse_non_interacting}

Using the effective action, we can compute observables such as the charge response function. To do so, let us first find the propagator of the bosonic field $\phi$. Since we consider the action up to quadratic order in $\phi$, we can find, in principle, the exact propagator of the Gaussian action by taking its inverse. However, because the interaction part $S^{(2)_{\text{int}}}$ of the action corresponds to a dense infinite dimensional matrix in terms of $\omega$ and $\theta$, obtaining the exact propagator is problematic. For that reason, we compute correlation functions by considering the interaction part perturbatively in reference to the non-interacting action. The bare propagator associated with the non-interacting action~\eqref{eq:disc_quad_bare} is 
\begin{widetext}
\begin{align}
G_{0}(\Omega,\mathbf{q};\omega,\theta;\omega',\theta')={}&-\frac{(2\pi)^{2}}{N(0)}\frac{\delta(\omega'-\omega)\delta(\theta'-\theta)}{(f_{0}(\omega+\Omega)-f_{0}(\omega))(\Omega-v_{F}(\mathbf{n}_{\theta}\cdot\mathbf{q}))}.\label{eq:non-int_prop}
\end{align}
\end{widetext}

\begin{figure}[ht]
    \centering
    \subfigure[]{\includegraphics[scale=0.9]{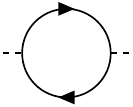}\label{fig:Pi0}}\hfill
    \subfigure[]{\includegraphics[scale=0.9]{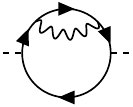}\label{fig:Pi1}}\hfill
    \subfigure[]{\includegraphics[scale=0.9]{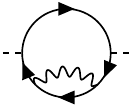}\label{fig:Pi2}}\hfill
    \subfigure[]{\includegraphics[scale=0.9]{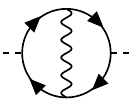}\label{fig:Pi3}}
    \caption{Feynmann diagrams for two-point density correlation function.}
    \label{fig:self}
\end{figure}

To compute the density correlation function, the density operators must be expressed in terms of $\phi$ as in Eq.~\eqref{eq:def_density_operator}. The deviation of the density operator $\delta\rho$ is at leading order given by (see Appendix~\ref{app:density_and_algebra})
\begin{align}
    \delta\rho(t,\mathbf{x})
    \approx{}\int\frac{d\omega d^{d-1}\theta}{(2\pi)^d} (\partial_{\omega} g_{0})\dot{\phi}, \label{eq:density_fluctuation_first_order}
\end{align}
and its Fourier transform is
\begin{align}
    \delta\hat{\rho}(\Omega,\mathbf{q})\approx i\int_{\omega\theta}(\partial_{\omega}g_{0})\Omega\hat{\phi}
    \approx i\int_{\omega\theta}(g_{0}(\omega+\Omega)-g_{0}(\omega))\hat{\phi}. \label{eq:den_op}
\end{align}
By using Eqs.~\eqref{eq:non-int_prop} and \eqref{eq:den_op}, the two-point density correlation function is
\begin{align}
\braket{\delta\hat{\rho}\,\delta\hat{\rho}}^{(0)}(\Omega,\mathbf{q})={}&N(0)\int\frac{d\omega d^{d-1}\theta}{(2\pi)^d} \frac{(f_{0}(\omega+\Omega)-f_{0}(\omega))}{\Omega-v_{F}q\cos\theta},\label{eq:non-int_den_correl}
\end{align}
where $\cos\theta=\mathbf{n}_{\theta}\cdot\hat{q}$. This is consistent with the particle-hole bubble contribution depicted in Fig.~\ref{fig:Pi0} at small momentum transfer in the equilibrium theory of a Fermi surface with the integration over $\xi_{\mathbf{p}}$ performed first (see Appendix~\ref{app:diagrammatic_calculation_charge_response}). Therefore, our effective Gaussian action in the non-interacting limit $S^{(2)_0}$ leads to the usual characteristic Fermi liquid behavior.

\subsection{Two-point density correlation function in the perturbative interacting limit} \label{subsec:charge_reponse_interactin_first_order}

We now consider the perturbative correction to the density correlation function due to the interacting part of the quadratic action. From Eq.~\eqref{eq:disc_quad_int}, the bilinear interaction vertices are given by
\begin{widetext}
\begin{subequations}\label{eq:interaction_vertices_quadratic_action}
\begin{align}
V_{\text{diag}}(\Omega,\mathbf{q};\omega,\theta;\omega',\theta')={}&-\frac{N(0)}{(2\pi)^{2}}(f_{0}(\omega+\Omega)-f_{0}(\omega))(\Sigma^{R}_{0}(\omega+\Omega)-\Sigma^{R}_{0}(\omega))\delta(\omega'-\omega)\delta(\theta'-\theta),\label{eq:interaction_vertices_quadratic_action_diag} \\
V_{\text{off}}(\Omega,\mathbf{q};\omega,\theta;\omega',\theta')={}&\frac{N(0)^{2}}{(2\pi)^{4}}(f_{0}(\omega+\Omega)-f_{0}(\omega))(f_{0}(\omega'+\Omega)-f_{0}(\omega'))D_{1}(\omega'-\omega,\theta'-\theta).\label{eq:interaction_vertices_quadratic_action_off}
\end{align}
\end{subequations}
Then, the leading order contributions to the two-point correlation function from the diagonal $V_{\text{diag}}$ and off-diagonal $V_{\text{off}}$ perturbations are respectively
\begin{subequations}\label{eq:perturb_charge_response_first_order}
\begin{align}
\braket{\delta\hat{\rho}\,\delta\hat{\rho}}^{(1)}_{\text{diag}}(\Omega,\mathbf{q})={}&N(0)\int_{\omega\theta}\frac{\Sigma^{R}_{0}(\omega+\Omega)-\Sigma^{R}_{0}(\omega)}{(\Omega-v_{F}q\cos\theta)^{2}}(f_{0}(\omega+\Omega)-f_{0}(\omega)),\label{eq:perturb_a_EFT}\\
\braket{\delta\hat{\rho}\,\delta\hat{\rho}}^{(1)}_{\text{off}}(\Omega,\mathbf{q})={}&-N(0)^{2}\int_{\omega_{1}\theta_{1}}\int_{\omega_{2}\theta_{2}}\frac{(f_{0}(\omega_{1}+\Omega)-f_{0}(\omega_{1}))(f_{0}(\omega_{2}+\Omega)-f_{0}(\omega_{2}))}{(\Omega-v_{F}q\cos\theta_{1})(\Omega-v_{F}q\cos\theta_{2})}D_{1}(\omega_{2}-\omega_{1},\theta_{2}-\theta_{1}).\label{eq:perturb_b_EFT}
\end{align}
\end{subequations}
These charge response corrections correspond to two-loop contributions in the equilibrium theory of a Fermi surface interacting with bosons through a Yukawa coupling. In particular, the first-order diagonal contribution $\braket{\delta\rho\,\delta\rho}^{(1)}_{\text{diag}}$ correspond to the self-energy corrections in Figs.~\ref{fig:Pi1} and~\ref{fig:Pi2}, whereas the off-diagonal term $\braket{\delta\rho\,\delta\rho}^{(1)}_{\text{off}}$ corresponds to the vertex correction of Fig.~\ref{fig:Pi3} (i.e., the so-called Maki-Thompson diagram). These correspondences are established by comparing with explicit calculations in Appendix~\ref{app:diagrammatic_calculation_charge_response}.
We can rewrite the total first-order correction as 
\begin{align}
\braket{\delta\hat{\rho}\,\delta\hat{\rho}}^{(1)}_{\text{diag}}+\braket{\delta\hat{\rho}\,\delta\hat{\rho}}^{(1)}_{\text{off}} ={}&N(0)^{2}\int_{\omega_{1}\theta_{1}}\frac{(f_{0}(\omega_{1}+\Omega)-f_{0}(\omega))}{(\Omega-v_{F}q\cos\theta_{1})}\notag\\
&\hspace{1cm}\times \int_{\omega_{2}\theta_{2}}\left[\frac{f_{0}(\omega_{2}+\Omega)-f_{0}(\omega_{2})}{(\Omega-v_{F}q\cos\theta_{1})}-\frac{f_{0}(\omega_{2}+\Omega)-f_{0}(\omega_{2})}{(\Omega-v_{F}q\cos\theta_{2})}\right]D_{1}(\omega_{2}-\omega_{1},\theta_{2}-\theta_{1}).
\end{align}
\end{widetext}
For $\theta_{2}=\theta_{1}$ (forward scattering) and small momentum transfer, this perturbative contribution vanishes, and the system exhibits the Fermi liquid-like behaviors, as discussed previously. Note that this does not happen for large momentum transfer.

Higher-order perturbative contributions to the charge response coming from the interacting part of the quadratic action $\braket{\delta\hat{\rho}\,\delta\hat{\rho}}^{(n)}$ are obtained by successive insertion of self-energy or simple vertex corrections shown in Fig.~\ref{fig:self}(b)-(d) to the bare particle-hole bubble, where these insertions do not cross or overlap. Diagrams obtained in this way encompass all contributions to the charge response for the quadratic action in the one-loop approximation.

\subsection{Angular momentum decomposition}\label{subsec:quadratic_angular}

Previous studies of the QBE for NFL in two dimensions showed that collective modes display very different behaviors~\cite {Kim1995}. In particular, for small angular momentum modes of the Fermi surface, there is a small-angle (forward-scattering) cancellation between the self-energy and the Landau interactions, which leads to the usual FL behavior. In comparison, large angular momentum modes show an anomalous singular behavior characteristic of NFL.

This subsection examines how this behavior transpires in our EFT by considering the angular momentum decomposition of the Gaussian action. Since we are working in two dimensions, we can introduce the following representation of $\hat{\phi}$,
\begin{align}
\hat{\phi}(\Omega,\mathbf{q};\omega,\theta)={}&\sum_{l}e^{i\theta l}\tilde{\phi}(\Omega,\mathbf{q};\omega,l),\label{eq:ang_decomp}
\end{align}
where $l$ is the angular momentum (i.e., the conjugate variable of $\theta$). Note that Eq.~\eqref{eq:ang_decomp} is valid for two dimensions, and one has to use spherical harmonics and generalizations thereof for the angular decomposition in three dimensions and higher. Then, the non-interacting part of the quadratic action in the angular momentum decomposition is given by 
\begin{widetext}
\begin{subequations}
\begin{align}\label{eq:ang_decomp_quad_bare}
S^{(2)_{0}}
={}&-\frac{N(0)}{2}\sum_{l}\int_{\Omega q\omega}F_{0}(\omega,\Omega)\tilde{\phi}(-\Omega,-\mathbf{q};\omega,l)(\Omega\tilde{\phi}(\Omega,\mathbf{q};\omega,l)-\frac{v_{F}q}{2}[\tilde{\phi}(\Omega,\mathbf{q},l+1)+\tilde{\phi}(\Omega,\mathbf{q};\omega,l-1)])
\end{align}
and the interacting part is
\begin{align}
S^{(2)_\text{int}}
={}&\frac{N(0)^{2}}{2}\sum_{l}\int_{\Omega q}\int_{\omega\omega'}\int_{\theta}F_{0}(\omega,\Omega)F_{0}(\omega',\Omega)D_{1}(\omega'-\omega,\theta)\tilde{\phi}(-\Omega,-\mathbf{q};\omega,l)[\tilde{\phi}(\Omega,\mathbf{q};\omega,l)-\cos(l\theta)\tilde{\phi}(\Omega,\mathbf{q};\omega',l)].
\end{align}
\end{subequations}
\end{widetext}
If we consider the scaling dimension of the free part of the action $S^{(2)_{0}}$ as zero, we can obtain the scaling dimension of the interacting part $S^{(2)_{\text{int}}}$.
We will attach the coupling constant $\mathcalligra{g}$ to $S^{(2)_{\text{int}}}$ and examine how it scales under the leading order renormalization group (RG). For concreteness, let us consider a case similar to Refs.~\cite{kim1994gauge, Kim1995} where a two-dimensional fermionic system interacts with a Landau damped massless boson (e.g., a critical boson or a transverse gauge field) whose propagator is of the form 
\begin{align}
    \mathcal{D}(\mathbf{q}, \omega)=\frac{1}{-i \gamma \frac{\omega}{|\mathbf{q}|}+\chi |\mathbf{q}|^\eta},
\end{align}
where $\gamma$ and $\chi$ are constants whereas $1<\eta\leq2$. Assuming $[\omega]=z$, we have $[\theta]=z/(\eta+1)$ and $[D_{1}]=-\eta[\theta]$. 
Then, we can define the critical angular momentum $l_{c}\propto k_{F}^{-1}\Omega^{1/(1+\eta)}$, and we can classify the behaviors depending on small $(l<l_{c})$ and large $(l>l_{c})$ angular momenta~\cite{Kim1995}.
For small angular momentum $l<l_{c}$ and $\omega=\omega'$, we can approximate $1-\cos(l\theta)\approx(l\theta)^{2}$ such that $[\mathcalligra{g}]=(d+1-\eta)[\theta]=-(\frac{3-\eta}{1+\eta})z$. So, the leading order RG flow equation for $g$ is 
\begin{align}
\frac{d\mathcalligra{g}}{d\ell}={}&-\frac{z(3-\eta)}{1+\eta}\mathcalligra{g},
\end{align}
which indicates that the interaction part becomes irrelevant, and the system behaves like a Fermi liquid. For large $l>l_{c}$, $\cos(l\theta)$ becomes highly oscillating as a function of $\theta$, so $[\mathcalligra{g}]=(\frac{\eta-1}{\eta+1})z$ and we have,
\begin{align}
\frac{d\mathcalligra{g}}{d\ell}=\frac{z(\eta-1)}{\eta+1}\mathcalligra{g}.
\end{align}
Then the interacting part becomes relevant when $\eta>1$ and we recover non-Fermi liquid behavior. Consequently, even within the quadratic Gaussian action, our effective action can capture the same physics exhibited by the quantum Boltzmann equation for non-Fermi liquids in the collisionless limit, such as the Fermi liquid behavior for $l<l_{c}$ collective modes (between $\Omega\propto q^{(1+\eta)/2}$ and $\Omega\propto q$) and non-Fermi liquid behavior for $l>l_{c}$ collective modes (below $\Omega\propto q^{(1+\eta)/2}$)~\cite{Kim1995}. 
We can apply same tree-level RG procedure in the angular momentum decomposition to analyze the higher-order action (see Appendix~\ref{app:higher}).
\section{Notes on the quadratic theory at all loop orders}\label{sec:general}

In the previous section, we restricted ourselves to the one-loop approximation in the self-energy expansion, $\Sigma^{R}=\Sigma^{R(1)}$. However, it is crucial to verify whether physical results of note, such as the forward scattering cancelation and separation of the angular momentum modes into rough and soft fluctuations, remain valid at higher order in the self-energy expansion or are simply artifacts of the lowest order approximation. To clarify these points, we derive and analyze the Gaussian action in $\phi$ for the general case where $\Sigma^{R}[g]=\sum_{n=1}\Sigma^{R(n)}[g]$.

\subsection{Derivation of the gaussian action at all loop orders} \label{subsec:gaussian_action_n_loop_order}

Following the same logic as in Sec.~\ref{subsec:quadatic_gaussian}, but now with the full functional expansion of the self-energy, we can write the action in powers of the $\phi$ field.
As shown in Appendix~\ref{app:general_higher}, the $\phi^{0}$ and $\phi^{1}$ order terms again correspond to a constant and a total time derivative, respectively. The quadratic action $S^{(2)}$ can still be split into $S^{(2)}=S^{(2)_{0}}+S^{(2)_{\text{int}}}$, where the non-interacting part $S^{(2)_{0}}$ is the same as in the one-loop approximation (see Eq.~\eqref{eq:cont_quad_bare}). In contrast, the interacting part of the Gaussian action is now 
\begin{align} 
S^{(2)_{\text{int}}}={}&\frac{1}{2}\int_{tx\omega\theta}(\partial_{\omega}g_{0})\dot{\phi}[(\partial_{\omega}\Sigma^{R}_{0})\dot{\phi}-\Delta^{(1)}_{\phi}\Sigma^{R}],\label{eq:general_cont_quad_int}
\end{align}
where $\Sigma^{R}_{0}=\sum_{n=1}\Sigma^{R(n)}[g_{0}]$. 
In the general case where the full expansion of the self-energy is considered, $\Delta_{\phi}^{(m)}\Sigma^{R(n)}$ can be written in terms of $\bar{\Delta}_{\phi}^{\{m_{i}\}}\Sigma^{R(n)}$, which is defined as
\begin{align}
&\bar{\Delta}_{\phi}^{(m_{1},\ldots,m_{i})}\Sigma^{R(n)}\notag\\&=\int_{\omega_{1}\theta_{1}}\ldots\int_{\omega_{n}\theta_{n}}D_{n}\Delta_{\phi}^{(m_{1})}f_{1}\ldots\Delta_{\phi}^{(m_{i})}g_{i}\prod_{j=i+1}^{n}g_{j}.
\end{align}
$\{m_{j}\}$ is a partition of $m$ which satisfies $\sum_{j=1}^{i}m_{j}=m$. For example, $\Delta_{\phi}^{(2)}\Sigma^{R(n)}=n\bar{\Delta}_{\phi}^{(2)}\Sigma^{R(n)}+\binom{n}{2}\bar{\Delta}_{\phi}^{(1,1)}\Sigma^{R(n)}$.
In Eq.~\eqref{eq:general_cont_quad_int}, $\Delta_{\phi}^{(1)}\Sigma^{R(n)}=n\bar{\Delta}_{\phi}^{(1)}\Sigma^{R(n)}$ such that  $\Delta_{\phi}^{(1)}\Sigma^{R}=\sum_{n=1}n\bar{\Delta}_{\phi}^{(1)}\Sigma^{R(n)}$. Accordingly, $\Sigma^{R}_{0}$ and $\Delta_{\phi}^{(1)}\Sigma^{R}$ now include contributions at all order in the generalized distribution $g$ that are absent from Eq.~\eqref{eq:disc_quad_int}.

In terms of the Fourier transformed field $\hat{\phi}(\omega,\theta;\Omega,\mathbf{q})$, the interacting part of the quadratic action takes the general form (after discretization of the frequency derivative)
\begin{widetext}
\begin{subequations}
\begin{align}
   S^{(2)_\text{int}}
={}&\frac{N(0)}{2}\int_{\Omega q\omega\theta}F_{0}(\omega,\Omega)(\Sigma^{R}_{0}(\omega+\Omega)-\Sigma^{R}_{0}(\omega))\hat{\phi}(\Omega,\mathbf{q};\omega,\theta)\hat{\phi}(-\Omega,-\mathbf{q};\omega,\theta)\notag\\
&-\sum_{n=1}\frac{n}{2}N(0)^{2}\int_{\Omega q\omega\theta}\int_{\omega'\theta'}\ldots\int_{\omega_{n-1}\theta_{n-1}}F_{0}(\omega,\Omega)F_{0}(\omega',\Omega)(\prod_{i=1}^{n-1}g_{0}(\omega_{i}))D_{n}(\omega_{1}-\omega,\theta_{1}-\theta;\ldots;\omega'-\omega,\theta'-\theta)\notag\\
&\qquad\qquad\qquad\qquad\qquad\qquad\qquad\times\hat{\phi}(\Omega,\mathbf{q};\omega,\theta)\hat{\phi}(-\Omega,-\mathbf{q};\omega',\theta') \label{eq:general_disc_quad_int}
\end{align}
Using the generic property $D_{n}(\omega_{1},\theta_{1},\ldots;\omega,\theta)=D_{n}(\omega_{1}-\omega,\theta_{1}-\theta;\ldots;\omega_{n}-\omega,\theta_{n}-\theta)$ for the generalized Landau functions,
 $S^{(2)_{\text{int}}}$ can be rewritten as
\begin{align}\label{eq:general_disc_quad_int_symm}
S^{(2)_{\text{int}}}={}&\sum_{n=1}\frac{n}{2}N(0)^{2}\int_{\Omega q\omega\theta}\int_{\omega'\theta'}\ldots\int_{\omega_{n-1}\theta_{n-1}}F_{0}(\omega,\Omega)F_{0}(\omega',\Omega)(\prod_{i=1}^{n-1}g_{0}(\omega_{i}))D_{n}(\omega_{1}-\omega,\theta_{1}-\theta;\ldots;\omega'-\omega,\theta'-\theta)\notag\\
&\qquad\qquad\qquad\qquad\qquad\qquad\qquad\times\hat{\phi}(\Omega,\mathbf{q};\omega,\theta)[\hat{\phi}(-\Omega,-\mathbf{q};\omega,\theta)-\hat{\phi}(-\Omega,-\mathbf{q};\omega',\theta')].
\end{align}
\end{subequations}
\end{widetext}
It can then be noted that forward scattering cancellation holds beyond the one-loop approximation since the interacting part of the quadratic action still vanishes near $\omega=\omega'$ and $\theta=\theta'$. Of course, one would expect the forward scattering cancellation to hold at all orders since it can usually be attributed to conservation laws~\cite{kim1994gauge, Kim1995}.

\subsection{Two-point density correlation function at all loop orders} \label{subsec:charge_reponse_n_loop_order}

The charge response can be computed perturbatively in the interacting part of the quadratic action using the Gaussian action derived above. The bilinear interaction vertices can once again be split into diagonal and off-diagonal parts as in Eq.~\eqref{eq:interaction_vertices_quadratic_action}. The diagonal interaction vertex is the same as in Eq.~\eqref{eq:interaction_vertices_quadratic_action_diag}, but $\Sigma_0$ is now expanded at all loop orders. The off-diagonal interaction vertex can be split into $V_{\text{off}}=\sum_{n} V_{\text{off},n}$, where $V_{\text{off},n}$ contains the contribution from $D_n$ and is given by 
\begin{widetext}
\begin{align}
    V_{\text{off},n}(\Omega,\mathbf{q};\omega,\theta;\omega',\theta')={}& \frac{n N(0)^{2}}{(2\pi)^{4}} \int_{\omega_{1}\theta_{1}}\ldots\int_{\omega_{n-1}\theta_{n-1}} F_{0}(\omega,\Omega)F_{0}(\omega',\Omega)(\prod_{i=1}^{n-1}g_{0}(\omega_{i}))D_{n}(\omega_{1}-\omega,\theta_{1}-\theta;\ldots;\omega'-\omega,\theta'-\theta). 
\end{align}
\end{widetext}
The charge response is then
\begin{align}
    \expval{\delta\hat{\rho}\delta\hat{\rho}} &= \expval{\delta\hat{\rho} \delta\hat{\rho}}^{(0)} + \expval{\delta\hat{\rho} \delta\hat{\rho}}^{(1)}_{\text{diag}} + \sum_{n=1}^{\infty} \expval{\delta\hat{\rho} \delta\hat{\rho}}^{(1)}_{\text{off,}n} + \ldots,
\end{align}
where the contribution from the non-interacting part of the action $\expval{\delta\hat{\rho} \delta\hat{\rho}}^{(0)}$ is the same as Eq.~\eqref{eq:non-int_den_correl} and corresponds to the bare particle-hole bubble. The first-order contribution from the diagonal term in the interacting part of the quadratic action $\expval{\delta\hat{\rho} \delta\hat{\rho}}^{(1)}_{\text{diag}}$ is the same as in Eq.~\eqref{eq:perturb_a_EFT}. However, $\Sigma_{0}^R$ is not restricted to the one-loop approximation anymore. For a Fermi surface Yukawa coupled to a boson, this term then corresponds to diagrams of the form Fig.~\ref{fig:self}(b)-(c), where the one-loop self-energy diagram is replaced by self-energy diagrams at all loop orders. Finally, the first perturbative correction to the charge response coming from the $n^{\text{th}}$ off-diagonal term is 
\begin{widetext}
\begin{align}
    \expval{\delta\hat{\rho} \delta\hat{\rho}}^{(1)}_{\text{off,}n} &= -n N(0)^2\int_{\theta_1 \omega_1 \theta_2 \omega_2} \frac{F_{0}(\omega_1,\Omega) F_{0}(\omega_2,\Omega) }{ (\Omega - v_{F}(\mathbf{n}_{\theta_1}\cdot\mathbf{q})) (\Omega - v_{F}(\mathbf{n}_{\theta_2}\cdot\mathbf{q})) } \nonumber \\
    &\quad \times \int_{\omega'_1 \theta'_1}\ldots\int_{\omega_{n-1}'\theta_{n-1}'}  \left(\prod_{i=1}^{n-1}g_{0}(\omega_{i}')\right) D_{n}(\omega_{1}'-\omega_1,\theta_{1}'-\theta_1;\ldots;\omega_2-\omega_1,\theta_2-\theta_1). \label{eq:charge_response_off_n_loop_order} 
\end{align}
\end{widetext}
These are higher-order vertex corrections to the charge response (see Appendix~\ref{app:off_diagoanl_AL}). Their exact correspondence with conventional diagrams depends on the particular theory under consideration and the associated relations between $D_{n}$ and $D_{1}$ (if any). For instance, in a theory of a Fermi surface with a Yukawa coupling to bosons, contributions from $D_{3}$ should yield the conventional Aslamazov-Larkin diagrams as argued in Appendix~\ref{app:off_diagoanl_AL}.

\subsection{Angular momentum decomposition at all loop orders} \label{subsec:angular_momentum_decomposition_n_loop_order}

Similarly, the angular momentum decomposition can be investigated in the general case for the self-energy expansion using the quadratic action \eqref{eq:general_disc_quad_int_symm}. 
Assuming we are working in two dimensions, we can get the angular momentum representation of the general quadratic theory by using Eq.~\eqref{eq:ang_decomp}. 
The non-interacting part of the Gaussian action is once again given by Eq.~\eqref{eq:ang_decomp_quad_bare}, whereas the interacting part is
\begin{widetext}
\begin{align}
S^{(2)_\text{int}}
={}&\sum_{n=1}\frac{n}{2}\sum_{l}N(0)^{2}\int_{\Omega q}\int_{\omega\omega'}\int_{\delta\theta'}\ldots\int_{\omega_{n-1}\delta\theta_{n-1}}F_{0}(\omega,\Omega)F_{0}(\omega',\Omega)(\prod_{i=1}^{n-1}g_{0}(\omega_{i}))D_{n}(\omega_{1}-\omega,\delta\theta_{1};\ldots;\omega'-\omega,\delta\theta')\notag\\
&\qquad\qquad\qquad\qquad\times\tilde{\phi}(-\Omega,-\mathbf{q};\omega,l)[\tilde{\phi}(\Omega,\mathbf{q};\omega,l)-\cos(l\theta)\tilde{\phi}(\Omega,\mathbf{q};\omega',l)],
\end{align}
\end{widetext}
where $\delta\theta'=|\theta'-\theta|$ and $\delta\theta_{i}=|\theta_{i}-\theta|$. For small angular momentum $l$, $S^{(2)_\text{int}}$ vanishes, and the system thus exhibits Fermi liquid behavior. On the other hand, the cancellation does not happen for the large angular momentum, and the interacting part of the quadratic action thus leads to the non-Fermi liquid-like rough fluctuations. This is analogous to the case discussed in Sec.~\ref{subsec:quadratic_angular} within the one-loop approximation. Therefore, the conclusions regarding the angular momentum decomposition of the EFT obtained in the one-loop approximation also hold in the general case.

\section{Discussion}\label{sec:discussion}
This work proposed a bosonized effective field theory for non-Fermi liquid. In our scheme, the fluctuating bosonic field $\phi$ parametrizes the Fermi surface deformation and may be interpreted as frequency-dependent local chemical potential variations in momentum space. The proposed action reproduces the quantum Boltzmann equation in the collisionless limit as its equation of motion. We obtained the Gaussian theory by expanding the non-linear action at quadratic order for $\phi$. The Gaussian action shows the usual Fermi liquid behavior for forward scattering due to a cancellation between the self-energy and vertex correction as seen from the Ward identity~\cite{kim1994gauge, Kim1995}. We computed the Gaussian contribution to the two-point density correlation function perturbatively and described how they relate to the loop corrections in the equilibrium theory of a Fermi surface interacting with gapless bosons through a Yukawa coupling. In the angular momentum decomposition of the Gaussian action, we demonstrated that it exhibits Fermi liquid-like behavior for small angular momentum and non-Fermi liquid behavior for large angular momentum. These results obtained only at the quadratic level are consistent with the previous studies of the quantum Boltzmann equation for non-Fermi liquid in the collisionless limit~\cite{Kim1995, Mandal2022}.
We have further demonstrated that these results are valid whether the self-energy is expanded at the one-loop order or considered in its most general functional expansion form.

A key point of our formalism is the use of the generalized distribution $f(\omega,\theta;t,\mathbf{x})=-i \int \frac{d \xi_\mathbf{p}}{2 \pi} G^{<}(\mathbf{p}, \omega ; \mathbf{x}, t)$ rather than the semiclassical phase space distribution used to describe the dynamics of ordinary Fermi liquids. Because of its dependence on the relative frequency $\omega$, the fluctuations of $f$ need to be parametrized by the generalized canonical transformation~\eqref{eq:canonical_transformatioon_quasiclassical_distribution} where the generalized Poisson bracket, with its additional time and frequency derivatives, is used. This novel approach allows us to effectively capture the intricate frequency dependence of the self-energy, which serves as a fundamental characteristic of non-Fermi liquid behavior. This explicit consideration of the frequency through the use of the generalized distributions and consideration of the self-energy is a crucial difference from the previous literature.

In this article, we have primarily focused on the quadratic action. Even at the Gaussian level, we have shown that our EFT reproduces most of the physics of the quantum Boltzmann equation, such as Fermi liquid and non-Fermi liquid behaviors in certain limits, which are consistent with previous results in the literature~\cite{Kim1995,Mandal2022}. Although much of the known physics can be reproduced at the quadratic order in $\phi$, higher-order effects in $\phi$ might still play an essential role in understanding physical observations. Our formalism provides an explicit way to consider these higher-order corrections systematically (see Appendices~\ref{app:higher} and~\ref{app:general_higher}). Non-linear effects and non-linear responses such as $n$-point density correlation function~\cite{Son2022} should thus be carefully considered in future studies.

When studying the proposed EFT, it was noted that the self-energy contribution to the action can be rewritten in terms of a functional $\Phi[f]$ (see Eq.~\eqref{eq:definition_generalized_luttinger-Ward}) that satisfies $\delta\Phi[g]/\delta f=\Sigma^{R}[g]$ --- in analogy with the conventional Luttinger-Ward functional. This intriguing observation requires clarification on whether the correspondence is fundamental or a mere coincidence. We expect further investigations of the functional to unveil a more concrete relationship between $\Phi[g]$ and the Luttinger-Ward functional or potentially other generating functionals of the equilibrium theory studied previously.

An alternative approach to the study of NFLs using our formalism would be to explicitly couple the effective action describing a Fermi surface to a critical bosonic mode. In this procedure, a non-interacting Fermi surface described by the effective action 
\begin{align}
    \tilde{S}_{\text{FS}}[\phi] &= \expval{\tilde{g}_0,U^{-1}\partial_t U} - \expval{\tilde{g},\xi_{\mathbf{p}}}
\end{align}
is first considered and coupled to a bosonic field $\Phi$
\begin{align}
    \tilde{S}_{\text{B}}[\Phi] &= \int_{t,\mathbf{x}} \left( \frac{1}{2} (\nabla \Phi(t,\mathbf{x}))^2+\frac{1}{2} k_0^2 \Phi(t,\mathbf{x})^2 \right), \label{eq:action_boson}
\end{align}
through, e.g., a term of the form  
\begin{align}
    \tilde{S}_{\text{int}}[\Phi, \phi] &= \lambda \int_{t,\mathbf{x}} \Phi(t,\mathbf{x})  \delta\rho(t,\mathbf{x}),
\end{align}
where $\delta\rho(t,\mathbf{x}) =  \int \frac{d\omega dp p^{d-1}}{(2\pi)^{d+1}} \left( U \tilde{g}_0 U^{-1} -  \tilde{g}_0\right)$. In such an approach, the phenomenological generalized Landau parameters $\tilde{D}_n$ do not appear. A similar approach has already been presented in Sec. IV.A of Ref.~\cite{Son2022} to show how one can recover a dynamical exponent $z=3$ anticipated from Landau damping at the quadratic order in the bosonized effective theory. The main difference here with Ref.~\cite{Son2022} would be that the above action retains the frequency dependence of the fermion distribution function, which we have argued to be crucial for NFL in Sec. II. Exactly how predictions by the two theories may differ needs to be clarified.

Our work further opens the doors to exciting and important generalizations. For instance, we have only considered the spinless Fermi surface. A natural extension will be the bosonization of the spinful (multiflavor) non-Fermi liquid. The quadratic action will be similar to the spinless case. Still, higher-order contributions to the action for $\phi$, such as the cubic terms, can differ significantly due to the nontrivial Lie algebra associated with the internal symmetry of the fermions~\cite{Son2022}. Therefore, analysis of higher-order terms in the action will become particularly important in studying spinful (multiflavor) Fermi surfaces.

\begin{acknowledgments}
We thank Daniel J. Schultz, Omid Tavakol, Hart Goldman, T.~Senthil, Yi-Hsien Du and D.~T.~Son for insightful discussions.  We acknowledge support from the Natural Sciences and Engineering Research Council of Canada (NSERC) and the Centre of Quantum Materials at the University of Toronto. F.D. is further supported by the Vanier Canada Graduate Scholarship. YBK is also supported by the Guggenheim Fellowship from the John Simon Guggenheim Memorial Foundation and the Simons Fellowship from the Simons Foundation. This work was partly performed at the Aspen Center for Physics, which is supported by National Science Foundation grant PHY-2210452. YBK acknowledges the support of the Advanced Study Group on ``Entanglement and Dynamics of Quantum Matter" at the Center for Theoretical Physics of Complex Systems in the Institute for Basic Science, where a part of the current work was done.
\end{acknowledgments}

\appendix

\section{Non-equilibrium field theory} \label{app:Non-equilibirum_field_theory}

To derive transport equations of metals, one has to consider a non-equilibrium situation. We here briefly review non-equilibrium field theory. The usual prescription is to work a closed-time Keldysh contour $\mathcal{C}$ (i.e., $\mathcal{C}$ describes a path from the distant past to the time of interest and back). The non-equilibrium Dyson's equation can be written as~\cite{kita2010introduction, rammer1986quantum, rammer2011quantum}
\begin{align} \label{eq:Dyson_equation_matrix_GF}
    \left(i  \frac{\partial}{\partial t_1} - H_0\right) \check{G}(1,2)-\int d3 \check{\Sigma}(1,3) \check{G}(3,2) = \mathds{1}\delta(1,2)
\end{align}
where $1\equiv (t_1, \mathbf{x}_1)$, $\delta(1,2)\equiv \delta(t_1-t_2)\delta^{(d)}(\mathbf{x}_1-\mathbf{x}_2)$,  $H_0$ is the non-interacting Hamiltonian, and the matrix Green's function and self-energy are
\begin{align}  \label{eq:Definition_matrix_GF_and_SE}
\check{G}=\left[\begin{array}{cc}G^{\mathcal{T}} & -G^{<} \\ G^{>} & -G^{\overline{\mathcal{T}}} \end{array}\right] \text{ and } \check{\Sigma}=\left[\begin{array}{ll} \Sigma^{\mathcal{T}} & -\Sigma^{<} \\ \Sigma^{>} & -\Sigma^{\overline{\mathcal{T}}} \end{array}\right]
\end{align}
with
\begin{subequations} \label{eq:Definition_of_all_GFs}
\begin{align} 
G^{>}\left(1, 2\right)= & -i\left\langle\psi\left(1\right) \psi^{\dagger}\left(2\right)\right\rangle, \\ 
G^{<}\left(1, 2\right)= & i\left\langle\psi^{\dagger}\left(2\right) \psi\left(1\right)\right\rangle, \\ 
G^{\mathcal{T}}\left(1, 2\right)= & \Theta\left(t_1-t_2\right) G^{>}\left(1, 2\right) \notag\\
&\hspace{1cm}+\Theta\left(t_2-t_1\right) G^{<}\left(1, 2\right), \\ 
G^{\overline{\mathcal{T}}}\left(1, 2\right)= & \Theta\left(t_2-t_1\right) G^{>}\left(1, 2\right) \notag\\
&\hspace{1cm}
 + \Theta\left(t_1-t_2\right) G^{<}\left(1, 2\right).
\end{align}
\end{subequations}

We note once again that Green's functions will not have spatial or temporal translation symmetry in the general non-equilibrium case. This implies that they will depend on both center of mass (i.e., $\mathbf{x}=(\mathbf{x}_1+\mathbf{x}_2)/2$ and $t=(t_1+t_2)/2$) and relative coordinates (i.e., $\mathbf{x}_{\mathrm{rel}}=\mathbf{x}_{2} - \mathbf{x}_1$ and $t_{\mathrm{rel}}=t_{2}-t_{1}$). To derive equations of motions, it is convenient to work in the Wigner representation
\begin{align}
    \check{G}\left(\omega, \mathbf{p};  t, \mathbf{x} \right) &= \int d^{d} x_{\text{rel}} dt_{\text{rel}} \check{G}(1,2) e^{-i\left(\mathbf{p} \cdot \mathbf{x}_{\text{rel}}- \omega t_{\text{rel}}\right)},
\end{align}
which is a Fourier transform with respect to the relative coordinates. To obtain the reformulation of Dyson's equation in this representation, one needs to apply the Wigner transformation to Eq.~\eqref{eq:Dyson_equation_matrix_GF} which contains matrix products. The Wigner transformation of a product of functions formally involves the Groenewold-Moyal product~\cite{moyal1949quantum, groenewold1946principles}
\begin{align}
    B * C =& B e^{\frac{i}{2}\left(\overleftarrow{\nabla}_{\mathbf{x}} \cdot \overrightarrow{\nabla}_{\mathbf{p}}-\overleftarrow{\partial}_t \overrightarrow{\partial}_{\omega}-\overleftarrow{\nabla}_{\mathbf{p}} \cdot \overrightarrow{\nabla}_{\mathbf{r}}+\overleftarrow{\partial}_{\omega} \overrightarrow{\partial}_t\right)} C,
\end{align}
which can be approximated at first order in the gradient expansion by
\begin{align}
    B*C \approx B C + \frac{i}{2} \left[B, C\right],
\end{align}
where we have introduced the generalized Poisson bracket defined in Eq.~\eqref{eq:generalized_Poisson_bracket}. Such a truncation is expected to be accurate when the microscopic length scale $l_m$ and time scale $t_m$ of interest are much smaller than the typical macroscopic spatial and temporal variation of the system (i.e., $l_m \gg 1/p_F$ and $t_m\gg 1/\varepsilon_F$)~\cite{kita2010introduction}. Equivalently stated, the first gradient approximation will be valid provided $\nabla_\mathbf{x}\cdot\nabla_{\mathbf{p}}\sim \abs{\nabla_{\mathbf{x}}}/p_F\ll 1$ and $\partial_{t} \partial_{\omega}\sim \abs{\partial_{t}}/\varepsilon_F\ll 1$, which implies that analysis should be restricted to particle-hole type excitations with typical net momentum and energy much less than the Fermi momentum $p_F$ and energy $\varepsilon_F$, respectively.

In the first-order gradient approximation of the Wigner representation, Dyson's equation leads to
\begin{equation} \label{eq:Dyson_equation_at_first_order_in_Wigner_representation}
    [G_{0}^{-1}-\Re\Sigma^{R}, G^{<}] - [\Sigma^{<}, \Re G^{R}] = G^{>} \Sigma^{>} - G^{<} \Sigma^{<},
\end{equation}
where we introduced $(G_0)^{-1} = \omega - \xi_{\mathbf{p}}$  with $\xi_{\mathbf{p}}=\epsilon(\mathbf{p})-\mu$ denoting the spectrum of $H_0$ (assumed to be independent of the center of mass position $\mathbf{x}$ and time $t$), the retarded Green's function 
\begin{equation}
    G^{R}(1,2) = \theta(t_1 - t_2) \left( G^{>}(1,2) - G^{<}(1,2) \right)
\end{equation}
and its self-energy $\Sigma^{R}$. The right-hand side of Eq.~\eqref{eq:Dyson_equation_at_first_order_in_Wigner_representation} is commonly referred to as the collision integral. 

\section{Useful relations}\label{app:useful}
In this section, we present derivations of some useful relations used in the main text.

\noindent \underline{1) Vanishing of $\int \frac{d\xi_{\mathbf{p}}}{2\pi}\text{Re}G^{R}$}
\begin{align}
    \int&\frac{d\xi_{\mathbf{p}}}{2\pi}\text{Re}G^{R}(\omega',\mathbf{p};t,\mathbf{x})\notag\\
    ={}&\int\frac{d\omega'}{2\pi}\mathcal{P}\frac{[1-f(\omega',\theta;t,\mathbf{x})]+f(\omega',\theta;t,\mathbf{x})}{\omega-\omega'}\notag\\
    ={}&\int\frac{d\omega'}{2\pi}\mathcal{P}\frac{1}{\omega-\omega'}=0,
\end{align}
where $\mathcal{P}$ means the principal value, and
\begin{align}
    \text{Re}&G^{R}(\omega',\mathbf{p};t,\mathbf{x})\notag\\
    ={}&-\int\frac{d\omega'}{2\pi}\mathcal{P}\frac{G^{>}(\omega',\mathbf{p};t,\mathbf{x})-G^{<}(\omega',\mathbf{p};t,\mathbf{x})}{\omega-\omega'},\\
    \int&\frac{d\xi_{\mathbf{p}}}{2\pi}[-iG^{<}(\omega,\mathbf{p};t,\mathbf{x})]={}f(\omega,\theta;t,\mathbf{x}),\\
    \int&\frac{d\xi_{\mathbf{p}}}{2\pi}[iG^{>}(\omega,\mathbf{p};t,\mathbf{x})]={}1-f(\omega,\theta;t,\mathbf{x}).
\end{align}

\noindent \underline{2) $\braket{\Delta_{\phi}^{(m)}\tilde{g},\Delta_{\phi}^{(n)}\Sigma^{R}} = \braket{\Delta_{\phi}^{(n)}\tilde{g},\Delta_{\phi}^{(m)}\Sigma^{R}}$}

\noindent Assuming an isotropic angular-resolved density of states $N(0,\theta)=N(0)$, such that $\int d\xi/(2\pi) N(0) \Delta^{(m)}_{\phi}\tilde{g} = \Delta^{(m)}_{\phi} g$, we have
\begin{align}
&\braket{\Delta_{\phi}^{(m)}\tilde{g},\Delta_{\phi}^{(n)}\Sigma^{R}}\notag\\
={}&\int_{t \mathbf{x} \omega \mathbf{p}}(\Delta_{\phi}^{(m)}\tilde{g})(\Delta_{\phi}^{(n)}\Sigma^{R})\notag\\
={}&\int_{t \mathbf{x} \omega \theta}\int_{\omega'\theta'}(\Delta_{\phi}^{(m)}g)D_{1}(\omega'-\omega,\theta'-\theta)(\Delta_{\phi}^{(n)}g')\notag\\
={}&\int_{t \mathbf{x} \omega \theta}\int_{\omega'\theta'}(\Delta_{\phi}^{(m)}g')D_{1}(\omega'-\omega,\theta'-\theta)(\Delta_{\phi}^{(n)}g)\notag\\
={}&\int_{t \mathbf{x} \omega \theta}(\Delta_{\phi}^{(n)}g)(\Delta_{\phi}^{(m)}\Sigma^{R})\notag\\
={}&\int_{t \mathbf{x} \omega \mathbf{p}}(\Delta_{\phi}^{(n)}\tilde{g})(\Delta_{\phi}^{(m)}\Sigma^{R})\notag\\
={}&\braket{\Delta_{\phi}^{(n)}\tilde{g},\Delta_{\phi}^{(m)}\Sigma^{R}},
\end{align}
where $\Delta_{\phi}^{(n)}\Sigma^{R}=\bar{\Delta}_{\phi}^{(n)}\Sigma^{R(1)}$ in the one-loop approximation (i.e., $\Sigma[g]=\Sigma^{(1)}[g]$) and we used the fact that $D_{1}$ does not change its sign under the change of variables~\eqref{eq:Change_of_variables_Landau_parameters}. Note that this identity is only valid in the one-loop approximation.

\section{Coadjoint orbit and other mathematical details} \label{app:math_details}

This appendix gives mathematical details regarding the underlying structure of the formalism. The main idea of this appendix is to adapt Sec.~II of Ref.~\cite{Son2022} by generalizing the group of canonical transformation $\mathcal{G}_{\text{s.c.}}$, its Lie algebra $\mathfrak{g}_{\text{s.c.}}$ and its adjoint and coadjoint representations to generalized canonical transformations $\mathcal{G}$ that have an additional dependence on center-of-mass time $t$ and frequency $\omega$ (i.e. the Fourier transform of the relative time). The mathematical details can be found in standard textbooks~\cite{kirillov2004lectures, isham1999modern, nakahara2018geometry} and are only provided here to make the article as self-contained as possible. 

\subsection{Lie algebra and its dual}

The Lie algebra of generalized canonical $\mathfrak{g}$ transformation is isomorphic to the space of smooth functions over $(t, \mathbf{x}; \omega, \mathbf{p})$ that vanish at infinity in all coordinate directions. This vector space is equipped with the generalized Poisson bracket~\eqref{eq:generalized_Poisson_bracket} as its Lie bracket. This space should be thought of as the space of observables, and, following the convention of Ref.~\cite{Son2022}, its elements are denoted with capital letters $F\in\mathfrak{g}$. 

The Lie algebra has a dual vector space $\mathfrak{g}^*$ defined as the space of maps from $\mathfrak{g}$ to the underlying field $\mathbb{R}$
\begin{align}
    f\in \mathfrak{g}^*: F\in \mathfrak{g} \mapsto \langle f,F \rangle\in\mathbb{R}.
\end{align}
The elements of the dual Lie algebra are denoted with lowercase letters. The inner product is defined as the integral over all coordinates
\begin{align}
    \langle f,F \rangle &= \int \frac{d\omega dt d^{d}x d^{d}p}{(2\pi)^{d+1}} f(t, \mathbf{x} ; \omega, \mathbf{p}) F(t, \mathbf{x} ; \omega, \mathbf{p})\notag\\
    &= \int_{t,\omega,\mathbf{x},\mathbf{p}} f(t, \mathbf{x} ; \omega, \mathbf{p}) F(t, \mathbf{x} ; \omega, \mathbf{p}).
\end{align}
We can interpret $f(t, \mathbf{x}; \omega, \mathbf{p})$ as a single-particle generalized distribution function and $\langle f,F \rangle$ as the expectation value of the observable $F$ for the distribution $f$. 

\subsection{Adjoint representation of the Lie algebra}

A Lie algebra representation on a vector space $V$ is a Lie algebra homomorphism 
\begin{align}
    \rho: \mathfrak{g} \rightarrow \operatorname{End}(V)
\end{align}
meaning that $\rho$ is a linear map that satisfies
\begin{align}
   \rho\left(\comm{F}{G}\right) = \rho(F)\circ\rho(G) - \rho(G)\circ\rho(F) \quad \forall F,G \in \mathfrak{g}
\end{align}
where $\circ$ denotes map composition. 

Since $\mathfrak{g}$ is itself a vector space, we can define the Lie algebra representation by its action on itself. This defines the adjoint representation
\begin{align}
    \ad{}&: \mathfrak{g} \rightarrow \operatorname{End}(\mathfrak{g}) \\
    \ad{}&: F \mapsto \ad{F} 
\end{align}
such that 
\begin{align}
   \ad{\comm{F}{G}} = \ad{F}\circ\ad{G} - \ad{G}\circ\ad{F}
\end{align}
In our case, we can define 
\begin{align}
    \ad{F}G = \comm{F}{G}.
\end{align}
With this definition, the above condition for the Lie algebra representation is easily verified to be respected by Jacobi's identity.

\subsection{Adjoint representation of the Lie group}

A group presentation on a vector space $V$ is a group homomorphism 
\begin{align}
    &R: \mathcal{G}\rightarrow \operatorname{GL}(V)\\
    &R(U)\circ R(V) = R(U\cdot V), \quad \forall U,V \in \mathcal{G},
\end{align}
where $\cdot$ denotes the group product in $\mathcal{G}$. Since the Lie algebra is a vector space, we can build a representation of the Lie group on its Lie algebra. This defines the adjoint representation of the Lie group: 
\begin{align}
    \Ad{}&: \mathcal{G} \rightarrow \operatorname{GL}(\mathfrak{g}) \\
    \Ad{}&: F \mapsto \Ad{F} 
\end{align}
which is a group homomorphism 
\begin{align}
   \Ad{F\cdot G} = \Ad{F}\circ\Ad{G}
\end{align}
defined by the adjoint representation of the Lie algebra as
\begin{align}
    \Ad{F} G = e^{\ad{F}} G  &= G + \comm{F}{G} +\frac{1}{2!} \comm{F}{\comm{F}{G}} + \ldots \notag\\
    &= e^{F} G e^{-F}.
\end{align}

\subsection{Coadjoint representation of the Lie algebra}

The coadjoint representation of the Lie algebra is a Lie algebra representation on $\mathfrak{g}^*$
\begin{align}
    \ad{}^*&: \mathfrak{g} \rightarrow \operatorname{End}(\mathfrak{g}^*) \\
    \ad{}^*&: F \mapsto \ad{F}^*
\end{align}
It has to respect
\begin{align}
   \Prod{\ad{G}^*f}{F} = -\Prod{f}{\ad{G}F}
\end{align}

In our case,
\begin{align}
    \ad{F}^* f = \comm{F}{f},
\end{align}
which can straightforwardly be shown to satisfy the above condition using integration by part.

\subsection{Coadjoint representation of the Lie group}

The coadjoint representation of the Lie group is
\begin{align}
    \Ad{}^*&: \mathcal{G} \rightarrow \operatorname{GL}(\mathfrak{g}^*) \\
    \Ad{}^*&: F \mapsto \Ad{F}^*,
\end{align}
such that 
\begin{align}
    \Ad{F}^* f = e^{\ad{F}^*} f  &= f + \comm{F}{f} +\frac{1}{2!} \comm{F}{\comm{F}{f}} + \ldots \notag\\
    &= e^{F} f e^{-F}.
\end{align}

The coadjoint representation of the Lie group has to respect 
\begin{align}
    \Prod{\Ad{G}F}{f} = \Prod{F}{\Ad{G^{-1}}f}.
\end{align}

\subsection{Group action, orbit and stabilizer}

A linear representation defines a group action on a vector space. For any group representation, one can define the orbit of a given vector $p$ as the set of all vectors that can be reached from $p$ through the group action. In the particular case of the coadjoint representation of a Lie group, the coadjoint orbit of a point $f_0\in\mathfrak{g}^*$ is then the space of all possible elements accessible from $f_0$
\begin{equation}
    \mathcal{O}_{f_0}=\left\{f \in \mathfrak{g}^{*} \mid \exists U \in \mathcal{G}: f=\operatorname{Ad}_U^* f_0\right\}.
\end{equation}
A key idea from the nonlinear bosonization framework put forth in Ref.~\cite{Son2022} and our current work is that all physically relevant valid distributions $f$ should be accessible by a generalized canonical transformation (i.e., time-evolution under the QBE) starting form a valid Fermi surface distribution $f_0$. $\mathcal{O}_{f_0}$ represents the space of physically relevant possible states, meaning that we can restrict our theory to $\mathcal{O}_{f_0}$ rather than $\mathfrak{g}^*$.

We note that on any group orbit, the group action is \emph{transitive}. This means that any two elements in the orbit can be connected by at least one element of the group. This statement trivially follows from the definition of the orbit. 

One can also define the stabilizer (or little group) on any element as the closed subgroup that acts trivially on $p$. In the case of the coadjoint orbit, the stabilizer of a given distribution $f_0\in\mathfrak{g}^*$ is 
\begin{equation}
    \mathcal{H}_{f_0}=\left\{U\in \mathcal{G} \mid  \operatorname{Ad}_U^* f_0 = f_0 \right\}.
\end{equation}

For transitive group action on any given set, there is an important theorem that states that the set on which the group acts has to be isomorphic to the set of all cosets of the stabilizer of any elements. Considering that the coadjoint representation acts transitively on the coadjoint orbit of any $f_0\in\mathfrak{g}^*$, the theorem implies that for any $f_0$ there is a bijection $j_{f_0}$ defined by
\begin{align}
j_{f_0}: \mathcal{G} / \mathcal{H}_{f_0} & \rightarrow \mathcal{O}_{f_0} \\
\Ad{U} \mathcal{H}_{f_0} & \mapsto \Ad{U} f_0.
\end{align}

In light of these formal statements, the gauge fixing procedure discussed in the main text consists of picking a single representative from every equivalence class $\mathcal{G}/\mathcal{H}_{f_0}$ in order to have a one-to-one correspondence between the representative and elements in $\mathcal{O}_{f_0}$. The above isomorphism is simply defined as $j_{f_0}(W)=\Ad{W} f_0$ for $W\in \mathcal{G}/\mathcal{H}_{f_0}$.

\subsection{Kirillov-Kostant-Souriau two-form and Wess-Zumino-Witten term}

Every coadjoint orbit is naturally equipped with a symplectic structure. On each orbit, there is a closed, non-degenerate, $G$-invariant 2-form $\omega_{\text{KKS}}$ referred to as the Kirillov-Kostant-Souriau (KKS) form~\cite{kirillov2004lectures}. Since $\mathfrak{g}^*$ is a vector space, it is isomorphic to its tangent space at all points $T_f\mathfrak{g}^{*} \simeq \mathfrak{g}^{*}$ $\forall f \in \mathfrak{g}^{*}$. The two-form can then be defined by its action on two distributions $g,k \in \mathfrak{g}^*$. It can be shown that the tangent vectors $g$ and $k$ at the point $f\in\mathcal{O}_{f_0}$ can be obtained by the coadjoint action of two Lie algebra elements $G, K \in \mathfrak{g}$: $\ad{G}^*f = g$ and $\ad{K}^*f = k$. We note that $G$ and $K$ here are not unique but rather representatives of an equivalence class. The KKS form is defined as 
\begin{align}
    \omega_{\text{KKS}}(g,k) = \Prod{f}{\comm{G}{K}}.
\end{align}

The effective action in our work and Ref.~\cite{Son2022} contains two parts: the Hamiltonian and the Wess-Zumino-Witten (WZW) terms. The Wess-Zumino-Witten term is defined with respect to the KKS form as 
\begin{align}
    S_{\text{WZW}}[\tilde{g}] = \int_0^{1} ds\; \omega_{\text{KKS}}\left( \partial_t \tilde{g}, \partial_s \tilde{g} \right),
\end{align}
where a new $s$ dimension is introduced and defined such that $\tilde{g}(s=1)=f$ and $\tilde{g}(s=0)=0$. As shown in Ref.~\cite{Son2022}, the WZW term can ultimately be rewritten as 
\begin{align}
    S_{\text{WZW}}[\tilde{g}] &= \Prod{\tilde{g}_0}{ U^{-1} \partial_t U }.
\end{align}.

\begin{widetext}
\section{Equation of motion}\label{app:general_action}

Let us consider the following self-energy contribution, which contains $n$ factors of $f$
\begin{align}
\tilde{\Sigma}^{R(n)}={}&\int_{1}\int_{2}\ldots\int_{n} \tilde{D}_{n}(\omega_{1}-\omega,\mathbf{p}_{1}-\mathbf{p};\ldots;\omega_{n}-\omega,\mathbf{p}_{n}-\mathbf{p}) \tilde{g}_{1} \tilde{g}_{2}\ldots \tilde{g}_{n},
\end{align}
where $\tilde{g}_{i}=\tilde{g}(t,\mathbf{x};\omega_{i},\mathbf{p}_{i})=U_{i} \tilde{g}_{0}(\omega_{i},\xi_{\mathbf{p}_{i}})U_{i}^{-1}$, $U_{i}=e^{-\phi(t,\mathbf{x};\omega_{i},\theta_{i})}$, $\int_{i}=\int_{\omega_{i}\mathbf{p}_{i}}$, and we assume that $\tilde{D}_{n}(\omega_{1}-\omega,\mathbf{p}_{1}-\mathbf{p};\ldots;\omega_{n}-\omega,\mathbf{p}_{n}-\mathbf{p}$ is invariant under the change of variables~\eqref{eq:Change_of_variables_Landau_parameters}. The self-energy changes under $U\rightarrow e^{-\alpha}U$, where $\alpha$ is an infinitesimal variation, according to
\begin{align}
 U^{-1}{}&\tilde{\Sigma}^{R(n)} U= U^{-1}\int_{1}\int_{2}\ldots\int_{n}( \tilde{D}_{n}\tilde{g}_{1}\tilde{g}_{2}\ldots \tilde{g}_{n})U\notag\\
={}& U^{-1}\int_{1}\int_{2}\ldots\int_{n}( \tilde{D}_{n}(U_{1}\tilde{g}_{0}U_{1}^{-1})(U_{2}\tilde{g}_{0}U_{2}^{-1})\ldots(U_{n}\tilde{g}_{0}U_{n}^{-1}))U\notag\\
\rightarrow& U^{-1}e^{\alpha}\int_{1}\int_{2}\ldots\int_{n}( \tilde{D}_{n}(e^{-\alpha_{1}}U_{1}\tilde{g}_{0}U_{1}^{-1}e^{\alpha_{1}})(e^{-\alpha_{2}}U_{2}\tilde{g}_{0}U_{2}^{-2}e^{\alpha_{2}})\ldots(e^{-\alpha_{n}}U_{n}\tilde{g}_{0}U_{n}^{-1}e^{\alpha_{n}}))e^{-\alpha}U\notag\\
=& U^{-1}\Bigl(\int_{1}\int_{2}\ldots\int_{n}\tilde{D}_{n}(e^{-\alpha_{1}}\tilde{g}_{1}e^{\alpha_{1}})(e^{-\alpha_{2}}\tilde{g}_{2}e^{\alpha_{2}})\ldots(e^{-\alpha_{n}}\tilde{g}_{n}e^{\alpha_{n}})\notag\\
&\qquad\qquad\qquad\qquad\qquad+[\alpha,\int_{1}\int_{2}\ldots\int_{n}\tilde{D}_{n}(e^{-\alpha_{1}}\tilde{g}_{1}e^{\alpha_{1}})(e^{-\alpha_{2}}\tilde{g}_{2}e^{\alpha_{2}})\ldots(e^{-\alpha_{n}}\tilde{g}_{n}e^{\alpha_{n}})]+\ldots\Bigr)U\notag\\
=& U^{-1}\Bigl(\int_{1}\int_{2}\ldots\int_{n}\tilde{D}_{n}(\prod_{i=1}^{n}\tilde{g}_{i}-n(\prod_{i=1}^{n-1}\tilde{g}_{i})[\alpha_{n},\tilde{g}_{n}]_{n}+\ldots)+[\alpha,\tilde{\Sigma}^{R(n)}+\ldots]+\ldots\Bigr)U,
\end{align}
where $[A,B]_{n}$ is the generalized Poisson bracket including the derivatives with respect to $\omega_{n}$ and $\theta_{n}$. It can then be noticed that for the contribution to the self-energy which contains $n$ generalized distribution function there is $n$ additional contributions to the linear order variation. 

Considering the general form of the action,
\begin{align}
    \tilde{S}=\braket{\tilde{g},\omega-\epsilon-\sum_{n=1}\tfrac{1}{n+1}\tilde{\Sigma}^{R(n)}[\tilde{g}]}.
\end{align}
we have
\begin{align}
\delta\tilde{S}={}& \braket{\tilde{g}_{0},-U^{-1}[\omega-\xi_{\mathbf{p}},\alpha]U-\frac{1}{n+1}U^{-1}\Bigl(n\int_{1}\int_{2}\ldots\int_{n}\tilde{D}_{n}(\prod_{i=1}^{n-1}\tilde{g}_{i})[\alpha_{n},\tilde{g}_{n}]_{n}-[\tilde{\Sigma}^{R(n)},\alpha]\Bigl)U}\notag\\
={}& \braket{[\omega-\xi_{\mathbf{p}}-\tfrac{1}{n+1}\tilde{\Sigma}^{R(n)},\tilde{g}],\alpha}-\sum_{n=1}\frac{n}{n+1} \braket{\int_{1}\int_{2}\ldots\int_{n}[\tilde{D}_{n} \tilde{g} (\prod_{i=1}^{n-1}\tilde{g}_{i}),\tilde{g}_{n}]_{n},\alpha_{n}}\notag\\
={}& \braket{[\omega-\xi_{\mathbf{p}}-\sum_{n=1}\tfrac{1}{n+1}\tilde{\Sigma}^{R(n)},\tilde{g}],\alpha}-\sum_{n=1}\frac{n}{n+1} \braket{\int_{1}\int_{2}\ldots\int_{n}[\tilde{D}_{n}(\prod_{i=1}^{n}\tilde{g}_{i}),\tilde{g}],\alpha}\notag\\
={}& \braket{[\omega-\xi_{\mathbf{p}}-\sum_{n=1}\tfrac{1}{n+1}\tilde{\Sigma}^{R(n)},\tilde{g}],\alpha}-\sum_{n=1}\frac{n}{n+1}\braket{[\tilde{\Sigma}^{R(n)},\tilde{g}],\alpha}\notag\\
={}& \braket{[\omega-\xi_{\mathbf{p}}-\tilde{\Sigma}^{R},\tilde{g}],\alpha},
\end{align}
so that the equation of motion is
\begin{align}
\int\frac{d\xi}{2\pi} N(0,\theta)[\omega-\xi-\tilde{\Sigma}^{R},\tilde{g}]=0
\end{align}
as mentioned in the main text. If $\tilde{D}_{n}$ is not invariant under the change of variables~\eqref{eq:Change_of_variables_Landau_parameters}, the coefficient of $\tilde{\Sigma}^{R(n)}$ in the effective action can differ.

\section{The generalized Luttinger-Ward functional \texorpdfstring{$\Phi[g]$}{}}\label{app:LW}

The self-energy contribution to our effective action including $n$ factors of the generalized distribution function can be written as
\begin{align}
\int_{t x \omega \theta}{}& g \sum_{n=1}\frac{1}{n+1}\Sigma^{R(n)} = \int_{t x \omega \theta} g \left( \frac{1}{2!}\int_{1}\frac{\delta\Sigma^{R}}{\delta g_1} g_1
+\frac{1}{3!}\int_{1}\int_{2}\frac{\delta^{2}\Sigma^{R}}{\delta g_1\delta g_2} g_1 g_2+\ldots\right) = \int_{t x} \Phi[f],
\end{align}
where we used $g_{i}=g(t,\mathbf{x};\omega_{i},\mathbf{p}_{i})$ and $\int_{i}=\int_{\omega_{i} \theta_{i}}$. Taking the derivative of the functional $\Phi[g]$ with respect to $g$, we have
\begin{align}
    \frac{\delta\Phi[g]}{\delta g}
    ={}&\int_{\omega_{1}\theta_{1}}\frac{\delta\Sigma^{R}}{\delta g_1}g_1 + \frac{1}{2!}\int_{\omega_{1}\theta_{1}}\int_{\omega_{1}\theta_{1}}\frac{\delta^{2}\Sigma^{R}}{\delta g_1\delta g_2 } f_1 f_2 + \ldots={}\Sigma^{R}[g].
\end{align}
So the functional has the property $\frac{\delta\Phi[g]}{\delta g}=\Sigma^{R}[g]$, which is analogous to the Luttinger-Ward functional. Then, following this analogy, we may consider the prefactor $1/(n+1)$ in front of the self-energy contribution in the effective action Eq.~\eqref{eq:action} as the inverse of the symmetry factor for the corresponding one-particle irreducible diagram including the $n$ generalized distribution functions. 
Thus, by using this generalized Luttinger-Ward functional, the effective action can be written as
\begin{align}
    S={}&\braket{\tilde{g},\omega-\xi}-\int_{tx}\Phi[g].
\end{align}

\section{Comments on non-linear corrections in the one-loop approximation}\label{app:higher}
We can also obtain higher-order terms in $\phi$ beyond the quadratic action, which gives us non-linear corrections.
From Eq.~\eqref{eq:effective_expand}, the general form of $n$-th order action is given by
\begin{align}
S^{(n)}={}&\braket{\Delta_{\phi}^{(n)}\tilde{g},\omega-\epsilon(p)-\tfrac{1}{2}\Sigma^{R}_{0}(\omega)}-\sum_{i=1}^{n}\braket{\Delta_{\phi}^{(n-i)}\tilde{g},\tfrac{1}{2}\Delta_{\phi}^{(i)}\Sigma^{R}}.
\end{align}

For example, the cubic action is
\begin{align}
    S^{(3)}={}&\braket{\Delta_{\phi}^{(3)}\tilde{g},\omega-\epsilon(p)-\Sigma^{R}_{0}(\omega)}-\braket{\Delta_{\phi}^{(2)}\tilde{g},\Delta_{\phi}^{(1)}\Sigma^{R}}.
\end{align}
The non-interacting part of this cubic term can be expressed in terms of $\phi$ as
\begin{align}
S^{(3)_{0}}={}&\braket{\Delta^{(3)}_{\phi}\tilde{g},\omega-\xi_{p}}\notag\\
={}&\frac{1}{6}\int_{tx\omega\theta}(\partial_{\omega}^{2}g_{0})\phi\dot{\phi}\ddot{\phi}
+\frac{1}{6}\int_{tx\omega\theta}(\partial_{\omega}g_{0})\dot{\phi}[p_{F}^{-1}\epsilon'(\mathbf{s}_{\theta}\cdot\nabla_{\mathbf{x}}\phi)^{2}+\epsilon''(\mathbf{n}_{\theta}\cdot\nabla_{\mathbf{x}}\phi)^{2}]\notag\\
&-\frac{1}{4}\int_{tx\omega\theta}\dot{\phi}^{2}(\partial_{\omega}g_{0})(\nabla_{\mathbf{p}}\epsilon)\cdot(\nabla_{\mathbf{x}}\partial_{\omega}\phi)\notag
\\
&-\frac{1}{6}\int_{tx\omega\theta}[(\nabla_{\mathbf{x}}\phi)\cdot(\nabla_{\mathbf{p}}\dot{\phi})-(\nabla_{\mathbf{p}}\phi)\cdot(\nabla_{\mathbf{x}}\dot{\phi})](\partial_{\omega}g_{0})(\nabla_{\mathbf{p}}\epsilon)\cdot(\nabla_{\mathbf{x}}\phi),\label{eq:cont_cubic_bare}
\end{align}
and the interacting part is
\begin{align}
S^{(3)_{\text{int}}}={}&-\braket{\Delta^{(3)}_{\phi}\tilde{g},\Sigma^{R}_{0}}-\braket{\Delta^{(2)}_{\phi}\tilde{g},\Delta^{(1)}_{\phi}\Sigma^{R}}\notag\\
={}&-\frac{1}{6}\int_{tx\omega\theta}[(\partial_{\omega}^{2}g_{0})(\partial_{\omega}\Sigma^{R}_{0})-(\partial_{\omega}g_{0})(\partial_{\omega}^{2}\Sigma^{R}_{0})]\phi\dot{\phi}\ddot{\phi}\notag\\
&+\frac{1}{2}\int_{tx\omega\theta}(\partial_{\omega}g_{0})\dot{\phi}[\dot{\phi}(\partial_{\omega}\Delta^{(1)}_{\phi}\Sigma^{R})-(\partial_{\omega}\phi)(\partial_{t}\Delta^{(1)}_{\phi}\Sigma^{R})]\notag\\
&+\frac{1}{2}\int_{tx\omega\theta}[(\nabla_{\mathbf{x}}\phi)\cdot(\nabla_{\mathbf{p}}\dot{\phi})-(\nabla_{\mathbf{p}}\phi)\cdot(\nabla_{\mathbf{x}}\dot{\phi})](\partial_{\omega}g_{0})\Delta^{(1)}_{\phi}\Sigma^{R}.\label{eq:cont_cubic_int}
\end{align}

The discretized non-interacting cubic action is
\begin{align}
S^{(3)_{0}}
={}&\frac{iN(0)}{6}\int_{\Omega q}'\int_{\omega\theta}[F_{0}(\omega,\Omega_{1})+F_{0}(\omega,-\Omega_{1})]\Omega_{2}\hat{\phi}_{1}\hat{\phi}_{2}\hat{\phi}_{3}\notag\\
&+\frac{iN(0)}{6}\int_{\Omega q}'\int_{\omega\theta}F_{0}(\omega,\Omega_{1})[p_{F}^{-1}\epsilon'(\mathbf{s}_{\theta}\cdot\mathbf{q}_{2})(\mathbf{s}_{\theta}\cdot\mathbf{q}_{3})+\epsilon''(\mathbf{n}_{\theta}\cdot\mathbf{q}_{2})(\mathbf{n}_{\theta}\cdot\mathbf{q}_{3})]\hat{\phi}_{1}\hat{\phi}_{2}\hat{\phi}_{3}\notag\\
&+\frac{iN(0)}{4}\int_{\Omega q}'\int_{\omega\theta}F_{0}(\omega,\Omega_{1})\hat{\phi}_{1}(v_{F}\mathbf{n}_{\theta}\cdot\mathbf{q}_{2})[\hat{\phi}_{2}(\omega+\Omega_{3})-\hat{\phi}_{2}(\omega)]\hat{\phi}_{3}\notag
\\
&-\frac{iN(0)}{6}\int_{\Omega q}'\int_{\omega\theta}F_{0}(\omega,\Omega_{1})[p_{F}^{-1}(\mathbf{q}_{2}\cdot\nabla_{\theta}\hat{\phi}_{1})\hat{\phi}_{2}-p_{F}^{-1}\hat{\phi}_{1}(\mathbf{q}_{1}\cdot\nabla_{\mathbf{\theta}}\hat{\phi}_{2})](v_{F}\mathbf{n}_{\theta}\cdot\mathbf{q}_{3})\hat{\phi}_{3},\label{eq:disc_cubic_bare}
\end{align}
and the discretized interacting cubic action is
\begin{align}
S^{(3)_{\text{int}}}={}&-\frac{iN(0)^{2}}{6}\int_{\Omega q}'\int_{\omega\theta}\int_{\omega'\theta'}[F_{0}(\omega',\Omega_{1})(F_{0}(\omega,\Omega_{2})+F_{0}(\omega,-\Omega_{2}))-F_{0}(\omega,\Omega_{1})(F_{0}(\omega',\Omega_{2})+F_{0}(\omega',-\Omega_{2}))]\notag\\
&\quad\quad\quad\quad\quad\quad\quad\quad\quad\quad\quad\quad\quad\times D_{1}(\omega'-\omega,\theta'-\theta)\hat{\phi}_{1}\hat{\phi}_{2}\hat{\phi}_{3}\notag\\
&+\frac{iN(0)^{2}}{2}\int_{\Omega q}'\int_{\omega\theta}\int_{\omega'\theta'}F_{0}(\omega,\Omega_{1})F_{0}(\omega',\Omega_{2})(D_{1}(\omega'-\omega-\Omega_{3},\theta'-\theta)-D_{1}(\omega'-\omega,\theta'-\theta))\hat{\phi}_{1}\hat{\phi}_{2}\hat{\phi}_{3}'\notag\\
&-\frac{iN(0)^{2}}{2}\int_{\Omega q}'\int_{\omega\theta}\int_{\omega'\theta'}F_{0}(\omega,\Omega_{1})F_{0}(\omega',\Omega_{3})D_{1}(\omega'-\omega,\theta'-\theta)\hat{\phi}_{1}(\hat{\phi}_{2}(\omega+\Omega_{3})-\hat{\phi}_{2}(\omega))\hat{\phi}_{3}'\notag\\
&-\frac{iN(0)^{2}}{2}\int_{\Omega q}'\int_{\omega\theta}\int_{\omega'\theta'}F_{0}(\omega,\Omega_{1})F_{0}(\omega',\Omega_{3})[p_{F}^{-1}(\mathbf{q}_{2}\cdot\nabla_{\theta}\hat{\phi}_{1})\hat{\phi}_{2}-p_{F}^{-1}\hat{\phi}_{1}(\mathbf{q}_{1}\cdot\nabla_{\theta}\hat{\phi}_{2})]D_{1}(\omega'-\omega,\theta'-\theta)\hat{\phi}_{3}'
\end{align}
where $\int_{\Omega q}'=\int\frac{d\Omega_{1} d^{d}q_{1}}{(2\pi)^{d+1}}\frac{d\Omega_{2} d^{d}q_{2}}{(2\pi)^{d+1}}\frac{d\Omega_{3} d^{d}q_{3}}{(2\pi)^{d+1}}\delta(\Omega_{1}+\Omega_{2}+\Omega_{3})\delta(\mathbf{q}_{1}+\mathbf{q}_{2}+\mathbf{q}_{3})$, and we used that $(\partial_{\omega}^{2}f_{0})\Omega^{2}\approx[f_{0}(\omega+\Omega)-f_{0}(\omega)]+[f_{0}(\omega-\Omega)-f_{0}(\omega)]$, $(\partial_{\omega}f_{0})\Omega\approx f_{0}(\omega+\Omega)-f_{0}(\omega)$, and $(\partial_{\omega}\hat{\phi}(\omega))\Omega\approx \hat{\phi}(\omega+\Omega)-\hat{\phi}(\omega)$. Note that we have also used the shorthand notation $F_{0}(\omega,\Omega)=f_{0}(\omega+\Omega)-f_{0}(\omega)$, $\hat{\phi}_{n}\equiv\hat{\phi}(\omega,\theta;\Omega_{n},\mathbf{q}_{n})$,  $\hat{\phi}'_{n}\equiv\hat{\phi}(\omega',\theta';\Omega_{n},\mathbf{q}_{n})$, and $\hat{\phi}(\omega)=\hat{\phi}(\omega,\theta;\Omega,\mathbf{q})$.

Similar to the discussion at the end of Sec.~\ref{subsec:quadratic_angular}, we can analyze the nonlinear terms in the higher-order actions by using the angular momentum decomposition. For example, in the cubic action in the example of the main text with $d=2$ and $z=1$, the angular momentum representation of the non-interacting cubic action is given by:
\begin{align}
S^{(3)_{0}}={}&\mathcalligra{g}_{30a}\frac{i}{6}\sum_{l_{1}l_{2}}N(0)\int_{\Omega q}'\int_{\omega}\Omega_{2}[F_{0}(\omega,\Omega_{1})+F_{0}(\omega,-\Omega_{1})]\tilde{\phi}_{1}(l_{1})\tilde{\phi}_{2}(l_{2})\tilde{\phi}_{3}(-l_{1}-l_{2})\notag\\
\notag&+\mathcalligra{g}_{30b}\frac{i}{4}\sum_{l_{1},l_{2}}N(0)\int_{\Omega q}'\int_{\omega}\frac{v_{F}q_{2}}{2}F_{0}(\omega,\Omega_{1})\tilde{\phi}_{1}(l_{1})[\tilde{\phi}_{2}(\omega+\Omega_{3},l_{2})-\tilde{\phi}_{2}(\omega,l_{2})][\tilde{\phi}_{3}(-l_{1}-l_{2}-1)e^{-i\vartheta_{3}}+\tilde{\phi}_{3}(-l_{1}-l_{2}+1)e^{i\vartheta_{3}}]\\
\notag&+\mathcalligra{g}_{30c}\frac{i}{6}\sum_{l_{1},l_{2}}N(0)\int_{\Omega q}'\int_{\omega}F_{0}(\omega,\Omega_{1})q_{2}q_{3}\frac{(p_{F}^{-1}\epsilon'+\epsilon'')}{2}\cos(\phi_{2}-\phi_{3})\tilde{\phi}_{3}(l_{3})\tilde{\phi}_{1}(l_{1})\tilde{\phi}_{2}(l_{2})\tilde{\phi}_{3}(-l_{1}-l_{2})\\
\notag&-\mathcalligra{g}_{30c}\frac{i}{6}\sum_{l_{1},l_{2}}N(0)\int_{\Omega q}'\int_{\omega}F_{0}(\omega,\Omega_{1})q_{2}q_{3}\frac{(p_{F}^{-1}\epsilon'-\epsilon'')}{2}[\tilde{\phi}_{3}(-l_{1}-l_{2}-2)e^{-i(\vartheta_{2}+\vartheta_{3})}+\tilde{\phi}_{3}(-l_{1}-l_{2}+2)e^{i(\vartheta_{2}+\vartheta_{3})})]\tilde{\phi}_{1}(l_{1})\tilde{\phi}_{2}(l_{2})\\
\notag&-\mathcalligra{g}_{30d}\frac{1}{6}\sum_{l_{1},l_{2}}N(0)\int_{\Omega q}'\int_{\omega}[F_{0}(\omega,\Omega_{1})-F_{0}(\omega,\Omega_{2})]v_{F}p_{F}^{-1}q_{2}q_{3}l_{1}\sin(\vartheta_{3}-\vartheta_{2})\tilde{\phi}_{1}(l_{1})\tilde{\phi}_{2}(l_{2})\tilde{\phi}_{3}(-l_{1}-l_{2})\\
&-\mathcalligra{g}_{30d}\frac{i}{6}\sum_{l_{1},l_{2}}N(0)\int_{\Omega q}'\int_{\omega}[F_{0}(\omega,\Omega_{1})-F_{0}(\omega,\Omega_{2})]v_{F}p_{F}^{-1}q_{2}q_{3}l_{1}\tilde{\phi}_{1}\tilde{\phi}_{2}[\tilde{\phi}_{3}(-l_{1}-l_{2}+2)e^{i(\vartheta_{2}+\vartheta_{3})}-\tilde{\phi}_{3}(-l_{1}-l_{2}-2)e^{-i(\vartheta_{2}+\vartheta_{3})}],
\end{align}
where $\vartheta_{i}$ is the angle for the momentum $\vec{q}_{i}$ and we attach the coupling constants $\mathcalligra{g}_{30\alpha}$ $(\alpha=a,b,c,d)$ for scaling analysis. The tree-level scaling dimensions for each coupling constant are $[\mathcalligra{g}_{30a,b}]=-1/2$ and $[\mathcalligra{g}_{30c,d}]=-3/2$, so they are irrelevant than the quadratic action.
The angular momentum representation of the interacting cubic action is given by
\begin{align}
S^{(3)_{\text{int}}}={}&-\frac{\mathcalligra{g}_{3ia}}{6}\sum_{l_{1},l_{2}}N(0)^{2}\int_{\Omega q}'\int_{\omega\omega'}\int_{\delta\theta'}[F_{0}(\omega',\Omega_{1})(F_{0}(\omega,\Omega_{2})+F_{0}(\omega,-\Omega_{2}))-F_{0}(\omega,\Omega_{1})(F_{0}(\omega',\Omega_{2})+F_{0}(\omega',-\Omega_{2}))]\notag\\&\quad\quad\quad\quad\quad\quad\quad\quad\quad\times D_{1}(\omega'-\omega,\delta\theta)\tilde{\phi}_{1}(l_{1})\tilde{\phi}_{2}(l_{2})\tilde{\phi}_{3}(-l_{1}-l_{2})\notag\\
\notag&+\mathcalligra{g}_{3ib}\frac{i}{2}\sum_{l_{1},l_{2}}N(0)^{2}\int_{\Omega q}'\int_{\omega\omega'}\int_{\delta\theta}F_{0}(\omega,\Omega_{1})F_{0}(\omega',\Omega_{2})(D_{1}(\omega'-\omega-\Omega_{3},\delta\theta)-D_{1}(\omega'-\omega,\delta\theta))\tilde{\phi}_{1}(l_{1})\tilde{\phi}_{2}(l_{2})\tilde{\phi}_{3}'(-l_{1}-l_{2})\\
&-\mathcalligra{g}_{3ic}\frac{i}{2}\sum_{l_{1},l_{2}}N(0)^{2}\int_{\Omega q}'\int_{\omega\omega}\int_{\delta\theta'}F_{0}(\omega,\Omega_{1})F_{0}(\omega',\Omega_{3})D_{1}(\omega'-\omega,\delta\theta)\tilde{\phi}_{1}(l_{1})(\tilde{\phi}_{2}(\omega+\Omega_{3},l_{2})-\tilde{\phi}_{2}(\omega,l_{2}))\tilde{\phi}_{3}'(-l_{1}-l_{2})\notag\\
&-\mathcalligra{g}_{3id}\frac{i}{2}\sum_{l_{1},l_{2}}N(0)^{2}\int_{\Omega q}'\int_{\omega\omega'}\int_{\delta\theta}p_{F}^{-1}q_{2}l_{1}[F_{0}(\omega,\Omega_{1})-F_{0}(\omega,\Omega_{2})]F_{0}(\omega',\Omega_{3})D_{1}(\omega'-\omega,\delta\theta)\tilde{\phi}_{1}(l_{1})\tilde{\phi}_{2}(l_{2})\notag\\&\quad\quad\quad\quad\quad\quad\quad\quad\quad\times [\tilde{\phi}_{3}'(-l_{1}-l_{2}+1)e^{i\vartheta_{3}}-\tilde{\phi}_{3}'(-l_{1}-l_{2}-1)e^{-i\vartheta_{3}}],
\end{align}
\end{widetext}
where  we again attach the coupling constants $\mathcalligra{g}_{3i\alpha}$ $(\alpha=a,b,c,d)$ for scaling analysis, similar to the non-interacting action. The tree-level scaling dimensions for each coupling constant are $[\mathcalligra{g}_{3i(a,b,c)}]=-\frac{1}{2}+\frac{\eta-1}{\eta+1}$ and $[\mathcalligra{g}_{3id}]=-\frac{3}{2}+\frac{\eta-1}{\eta+1}$. Then, $\mathcalligra{g}_{3id}$ is irrelevant than quadratic action regardless of value of $\eta$. $\mathcalligra{g}_{3i(a,b,c)}$ is irrelevant for $\eta<3$, but it can be relevant when $\eta>3$. However, as mentioned before, since we consider $1\leq\eta<2$, they will be irrelevant with the region which we are concerned.

\section{Quadratic action for \texorpdfstring{$\phi$ in the one-loop approximation}{}}\label{app:quadratic}
The Gaussian action in the one-loop approximation is given by
\begin{align}
    S^{(2)}={}&\braket{\Delta_{\phi}^{(2)}\tilde{g},\omega-\xi_{\mathbf{p}}-\Sigma^{R}_{0}}
    - \braket{\Delta_{\phi}^{(1)}\tilde{g},\tfrac{1}{2}\Delta_{\phi}^{(1)}\Sigma^{R}}.
\end{align}
It can be computed as follows. The first part is
\begin{align}
    &\braket{\Delta_{\phi}^{(2)}\tilde{g},\omega-\xi_{\mathbf{p}}-\Sigma^{R}_{0}}\notag\\
    ={}&\frac{1}{2}\braket{[\phi,[\phi,\tilde{g}_{0}]],\omega-\xi_{\mathbf{p}}-\Sigma^{R}_{0}}\notag\\
    ={}&-\frac{1}{2} \braket{[\phi, \tilde{g}],[\phi,\omega-\xi_{\mathbf{p}}-\Sigma^{R}_{0}]}\notag\\
    ={}&\frac{1}{2} \braket{(\partial_{\omega} \tilde{g}_0)\dot{\phi},-\dot{\phi}-(\nabla_{\mathbf{p}}\xi_{\mathbf{p}})\cdot(\nabla_{\mathbf{x}}\phi)+(\partial_{\omega}\Sigma^{R}_{0})\dot{\phi}]}\notag\\
    ={}&-\frac{1}{2}\int_{t\omega x \xi} N(0)(\partial_{\omega} \tilde{g}_{0})\dot{\phi}[\dot{\phi}+(\nabla_{\mathbf{p}}\xi_{\mathbf{p}})\cdot(\nabla_{\mathbf{x}}\phi)]\notag\\
    &+\frac{1}{2}\int_{t\omega x \xi} N(0) (\partial_{\omega} \tilde{g}_{0})\dot{\phi}^{2}(\partial_{\omega}\Sigma^{R}_{0})\notag\\
    ={}&-\frac{1}{2}\int_{t\omega x \theta} (\partial_{\omega}g_{0})\dot{\phi}[\dot{\phi}+(\nabla_{\mathbf{p}}\xi_{\mathbf{p}})\cdot(\nabla_{\mathbf{x}}\phi)]\notag\\
    &+\frac{1}{2}\int_{t\omega x \theta}(\partial_{\omega}g_{0})\dot{\phi}^{2}(\partial_{\omega}\Sigma^{R}_{0}),
\end{align}
and second part is
\begin{align}
    \braket{\Delta_{\phi}^{(1)}\tilde{g},\tfrac{1}{2}\Delta_{\phi}^{(1)}\Sigma^{R}} ={}&-\frac{1}{2}\braket{[\phi,\tilde{g}_{0}],\Delta_{\phi}^{(1)}\Sigma^{R}}\notag\\
    ={}&\frac{1}{2} \braket{(\partial_{\omega} \tilde{g}_{0})\dot{\phi},\Delta_{\phi}^{(1)}\Sigma^{R}}\notag\\
    ={}&\frac{1}{2}\int_{tx\omega\theta}(\partial_{\omega}g_{0})\dot{\phi}\Delta_{\phi}^{(1)}\Sigma^{R}.
\end{align}

Note that $\Sigma^{R}_{0}$ only depends on $\omega$.
Then, as a result, we can obtain Eq.~\eqref{eq:cont_quad_bare} and \eqref{eq:cont_quad_int} in the main text. If we perform the Fourier transformation for $\phi$, we can easily obtain Eq.~\eqref{eq:disc_quad_bare} in the main text. In the case of Eq.~\eqref{eq:disc_quad_int},
\begin{widetext}
\begin{align}
    S^{(2)_{\text{int}}}={}&\frac{1}{2}\int_{t\omega x \theta}(\partial_{\omega}g_{0})\dot{\phi}[(\partial_{\omega}\Sigma^{R}_{0})\dot{\phi}-\Delta_{\phi}^{(1)}\Sigma^{R}]\notag\\
    ={}&\frac{1}{2}\int_{t \omega x \theta}\int_{\omega'\theta'}(\partial_{\omega}g_{0})\dot{\phi}D_{1}(\omega'-\omega,\theta'-\theta)(\partial_{\omega'}g_{0}')\left[\dot{\phi}-\dot{\phi}'\right]\notag\\
    ={}&\frac{N(0)^{2}}{2}\int_{\Omega q\omega\theta}\int_{\omega'\theta'}[(\partial_{\omega}f_{0})\Omega][(\partial_{\omega'}f_{0}')\Omega]D_{1}(\omega'-\omega,\theta'-\theta)\hat{\phi}(\Omega,\mathbf{q};\omega,\theta)[\hat{\phi}(\Omega,\mathbf{q};\omega,\theta)-\hat{\phi}(\Omega,\mathbf{q};\omega',\theta')]\notag\\
    \approx{}&\frac{N(0)^{2}}{2}\int_{\Omega q\omega\theta}\int_{\omega'\theta'}[f_{0}(\omega+\Omega)-f_{0}(\omega)][f_{0}(\omega'+\Omega)-f_{0}(\omega')]D_{1}(\omega'-\omega,\theta'-\theta)\hat{\phi}(\Omega,\mathbf{q};\omega,\theta)\notag\\
    &\qquad\qquad\qquad\times[\hat{\phi}(-\Omega,-\mathbf{q};\omega,\theta)-\hat{\phi}(-\Omega,-\mathbf{q};\omega',\theta')]\notag\\
    ={}&\frac{N(0)^{2}}{2}\int_{\Omega q\omega\theta}\int_{\omega'\theta'}[f_{0}(\omega+\Omega)-f_{0}(\omega)][f_{0}(\omega'+\Omega)-f_{0}(\omega')]D_{1}(\omega'-\omega,\theta'-\theta)\hat{\phi}(\Omega,\mathbf{q};\omega,\theta)\hat{\phi}(-\Omega,-\mathbf{q};\omega,\theta)\notag\\
    &-\frac{N(0)^{2}}{2}\int_{\Omega q\omega\theta}\int_{\omega'\theta'}[f_{0}(\omega+\Omega)-f_{0}(\omega)][f_{0}(\omega'+\Omega)-f_{0}(\omega')]D_{1}(\omega'-\omega,\theta'-\theta)\hat{\phi}(\Omega,\mathbf{q};\omega,\theta)\hat{\phi}(-\Omega,-\mathbf{q};\omega',\theta')\notag\\
    ={}&\frac{N(0)}{2}\int_{\Omega q\omega\theta}[f_{0}(\omega+\Omega)-f_{0}(\omega)][\Sigma^{R}_{0}(\omega+\Omega)-\Sigma^{R}_{0}(\omega)]\hat{\phi}(\Omega,\mathbf{q};\omega,\theta)\hat{\phi}(-\Omega,-\mathbf{q};\omega,\theta)\notag\\
    &-\frac{N(0)^{2}}{2}\int_{\Omega q\omega\theta}\int_{\omega'\theta'}[f_{0}(\omega+\Omega)-f_{0}(\omega)][f_{0}(\omega'+\Omega)-f_{0}(\omega')]D_{1}(\omega'-\omega,\theta'-\theta)\hat{\phi}(\Omega,\mathbf{q};\omega,\theta)\hat{\phi}(-\Omega,-\mathbf{q};\omega',\theta'),
\end{align}
where we used $(\partial_{\omega}f_{0})\Omega\approx f_{0}(\omega+\Omega)-f_{0}(\omega)$.

\section{Derivation of the density operator and algebra density}\label{app:density_and_algebra} 

We can compute the density operator defined in Eq.~\eqref{eq:def_density_operator} order by order in $\phi$. At order $\phi^0$, the density operator in $d$-dimensions assuming an isotropic dispersion is
\begin{align}
    \rho^{(0)}(t,\mathbf{x},\theta) &= \int\frac{d\omega dp}{(2\pi)^{d+1}} p^{d-1} \tilde{g}_{0}(\omega,p) = \int\frac{d\omega dp}{(2\pi)^{d+1}} p^{d-1} 2\pi \delta\left(\omega - \xi_p \right) \Theta(-\omega) = \int\frac{dp}{(2\pi)^{d}} p^{d-1} \Theta\left(\xi_p \right) = \frac{p_F^{d}}{d (2\pi)^d}.
\end{align}
At this leading order, the density operator is the same as in the Fermi surface bosonization formalism of Ref.~\cite{Son2022}. At the next order in $\phi$, the density operator is
\begin{align}
    \rho^{(1)}(t,\mathbf{x},\theta) &= - \int\frac{d\omega dp p^{d-1}}{(2\pi)^{d+1}}  \left[ \phi, \tilde{g_0}\right]\nonumber \\
    &= - \int\frac{d\omega dp}{(2\pi)^{d+1}}  p^{d-1} \left( -\partial_t\phi \partial_{\omega} \tilde{g}_0 + \nabla_{\mathbf{x}}\phi \cdot \nabla_{\mathbf{p}}\tilde{g}_0 \right) \nonumber \\
    &= - \int\frac{d\omega dp}{(2\pi)^{d}}  p^{d-1} \left( -\partial_t\phi \partial_{\omega} \left( \delta\left(\omega - \xi_p \right) \Theta(-\omega) \right) + \nabla_{\mathbf{x}}\phi \cdot \mathbf{n}_{\theta} \partial_{p} \delta\left(\omega - \xi_p \right) \Theta(-\omega) \right) \nonumber \\
    &=  \int\frac{d\omega dp}{(2\pi)^{d}}   \left( -p^{d-1} \partial_t\partial_{\omega}\phi +  \partial_{p} \left( p^{d-1} \nabla_{\mathbf{x}}\phi \cdot \mathbf{n}_{\theta} \right)  \right) \delta\left(\omega - \xi_p \right) \Theta(-\omega) \nonumber \\
    &=  \int\frac{dp}{(2\pi)^{d}}   \left[ -p^{d-1} \partial_t\partial_{\omega}\phi + (d-1) p^{d-2} \nabla_{\mathbf{x}}\phi \cdot \mathbf{n}_{\theta} \right]_{\omega=\xi_p} \Theta\left(- \xi_p \right)  \nonumber \\
    &=  \int_{0}^{p_F}\frac{dp}{(2\pi)^{d}}   \left( -p^{d-1} \left[\partial_t\partial_{\omega}\phi\right]_{\omega=\xi_p} + (d-1) p^{d-2} \mathbf{n}_{\theta} \cdot \nabla_{\mathbf{x}}\left.\phi\right|_{_{\omega=\xi_p}}  \right).
\end{align}
Assuming the $\phi$ fields are frequency-independent, in which case the formalism essentially recovers the formulation of Ref.~\cite{Son2022}, the first-order density operator is 
\begin{align}
    \rho^{(1)}(t,\mathbf{x},\theta) &= \frac{p_F^{d-1}}{(2\pi)^d} \mathbf{n}_{\theta}\cdot \nabla_{\mathbf{x}}\phi(t,\mathbf{x},\theta),
\end{align}
which is the same as in~\cite{Son2022}. 

If we now fix the density of states to its value on the Fermi surface, the charge displacement from equilibrium is at leading order given by
\begin{align}
    \delta\rho(t,\mathbf{x}) &= \int\frac{dt_1 d\omega_1 d^{d}x_1 d^{d}p_1}{(2\pi)^{d+1}} \left( \tilde{g} -\tilde{g}_0 \right) \delta^{d}(\mathbf{x} - \mathbf{x}_1) \delta(t-t_1) \nonumber \\
    &\approx - \int\frac{dt_1 d\omega_1 d^{d}x_1 d^{d-1}\theta_1 d\xi_1}{(2\pi)^{d+1}} N(0) \comm{\phi}{\tilde{g}_0} \delta^{d}(\mathbf{x} - \mathbf{x}_1) \delta(t-t_1) + \ldots \nonumber \\
    &= - \int\frac{dt_1 d\omega_1 d^{d}x_1 d^{d-1}\theta_1 d\xi_1}{(2\pi)^{d+1}} N(0) \left(-\partial_{t_1}\phi \partial_{\omega_1}\tilde{g_0} + \nabla_{\mathbf{x}}\phi\cdot\nabla_{\mathbf{p}}\tilde{g}_0 \right) \delta^{d}(\mathbf{x} - \mathbf{x}_1) \delta(t-t_1) + \ldots \nonumber \\
    &= -\int \frac{d \omega d^{d-1}\theta d\xi}{(2\pi)^{d}} N(0) (\partial_{\omega}\partial_t\phi) \delta(\omega-\xi)\Theta(-\omega) + \ldots \nonumber \\
    &=  -\int \frac{d \omega d^{d-1}\theta}{(2\pi)^{d}} N(0) (\partial_{\omega}\partial_t\phi) \Theta(-\omega) + \ldots \nonumber \\
    &=  \int \frac{d \omega d^{d-1}\theta}{(2\pi)^{d}} N(0)  \partial_t\phi(t,\mathbf{x},\omega,\theta) \partial_{\omega} f_{0}(\omega) + \ldots
 \end{align}
This is the expression used in Eq.~\eqref{eq:density_fluctuation_first_order}

The commutator of the density operators introduced in Eq.~\eqref{eq:density_algebra} can be computed as follows
\begin{align}
&\comm{\rho[\phi](t_1,\mathbf{x}_1,\theta_1)}{\rho[\phi](t_2,\mathbf{x}_2,\theta_2)}_Q \nonumber\\
&= \left\langle \tilde{g}, i \comm{\rho(t_1,\mathbf{x}_1,\theta_1)}{\rho(t_2,\mathbf{x}_2,\theta_2)} \right\rangle \nonumber\\
&= i \int \frac{d\omega d t d^{d}x d^{d}p}{(2\pi)^{d+1}} \tilde{g}(t,\mathbf{x};\omega,\mathbf{p})  \comm{\rho(t_1,\mathbf{x}_1,\theta_1)}{\rho(t_2,\mathbf{x}_2,\theta_2)} \nonumber\\
&= -i \int \frac{d\omega d t d^{d}x d^{d}p}{(2\pi)^{d+1}} \comm{\tilde{g}(t,\mathbf{x};\omega,\mathbf{p})}{\rho(t_2,\mathbf{x}_2,\theta_2)} \rho(t_1,\mathbf{x}_1,\theta_1) \nonumber \\
&= -i \int \frac{d\omega d t d^{d}x d^{d}p}{(2\pi)^{d+1}} \comm{\tilde{g}_0(\omega,p)}{ \delta\left(t-t_2\right) \delta^{d}\left(\mathbf{x}-\mathbf{x}_2\right) \delta^{d-1}\left(\theta-\theta_2\right) } \delta\left(t-t_1\right) \delta^{d}\left(\mathbf{x}-\mathbf{x}_1\right) \delta^{d-1}\left(\theta-\theta_1\right) + \ldots \nonumber \\
&= -i \int \frac{d\omega d t d^{d}x d^{d}p}{(2\pi)^{d+1}} \left(\partial_{\omega}\tilde{g}_0(\omega,p) \partial_t \delta\left(t-t_2\right) \delta^{d}\left(\mathbf{x}-\mathbf{x}_2\right) \delta^{d-1}\left(\theta-\theta_2\right)\right.\nonumber \\
&\left. \hspace{2.5cm} -\nabla_{\mathbf{p}}\tilde{g}_0(\omega,p)\cdot\nabla_{\mathbf{x}} \delta^{d}\left(\mathbf{x}-\mathbf{x}_2\right) \delta\left(t-t_2\right) \delta^{d-1}\left(\theta-\theta_2\right) \right)\delta\left(t-t_1\right) \delta^{d}\left(\mathbf{x}-\mathbf{x}_1\right) \delta^{d-1}\left(\theta-\theta_1\right) + \ldots \nonumber \nonumber \\
&= -i \int \frac{d\omega dp p^{d-1}}{(2\pi)^{d}} \left(\partial_{\omega}\left(\delta(\omega-\xi_p)\Theta(-\omega) \right) \partial_{t_1}\delta\left(t_1-t_2\right)  \delta^{d-1}\left(\theta_1-\theta_2\right) \delta^{d}\left(\mathbf{x}_1 - \mathbf{x}_2\right) \right. \nonumber \\ 
&\hspace{4cm} \left. - \partial_p \delta(\omega-\xi_p)\Theta(-\omega) \mathbf{n}_{\theta}\cdot\nabla_{\mathbf{x}_1}\delta^{d}\left(\mathbf{x}_1-\mathbf{x}_2\right) \delta\left(t_1-t_2\right)  \delta^{d-1}\left(\theta_1-\theta_2\right) \right) + \ldots \nonumber \\
&= -i \delta\left(t_1-t_2\right)  \delta^{d-1}\left(\theta_1-\theta_2\right) \mathbf{n}_{\theta}\cdot\nabla_{\mathbf{x}_1}\delta^{d}\left(\mathbf{x}_1-\mathbf{x}_2\right) \int \frac{d\omega dp }{(2\pi)^{d}}  (d-1)p^{d-2} \delta(\omega-\xi_p)\Theta(-\omega) + \ldots  \nonumber \\
&= -i \delta\left(t_1-t_2\right) \mathbf{n}_{\theta}\cdot\nabla_{\mathbf{x}_1}\delta^{d}\left(\mathbf{x}_1-\mathbf{x}_2\right)  \delta^{d-1}\left(\theta_1-\theta_2\right) \int \frac{ dp }{(2\pi)^{d}}  (d-1)p^{d-2} \Theta(-\xi_p) + \ldots \nonumber \\
&= -i \frac{p_F^{d-1}}{(2\pi)^d} \delta\left(t_1-t_2\right) \mathbf{n}_{\theta}\cdot\nabla_{\mathbf{x}_1}\delta^{d}\left(\mathbf{x}_1-\mathbf{x}_2\right)  \delta^{d-1}\left(\theta_1-\theta_2\right) + \ldots 
\end{align}

\section{Diagrammatic calculation of the charge response}\label{app:diagrammatic_calculation_charge_response} 

In this appendix, we compute the diagrammatic contributions to the charge response for fermions with a spherical Fermi surface coupled to bosons through a Yukawa coupling (with the fermion-boson coupling set to unity). The goal of these calculations is to have a direct comparison with the charge response contributions in the bosonized theory and see which diagrams are produced at a given order in the charge response expansion.

\subsection{Bare particle-hole bubble} \label{app:diagrammatic_calculation_charge_response_bare}

The bare particle-hole bubble contribution depicted in Fig.~\ref{fig:Pi0} for small momentum transfer $\mathbf{q}$ in imaginary time is
\begin{align}
\Pi_{0}&(i\nu,\mathbf{q}) ={} -\int\frac{d\nu'}{2\pi}\frac{d^{2}p}{(2\pi)^{2}} \frac{1}{i\nu'+i\nu-\xi_{\mathbf{p}+\mathbf{q}}}\frac{1}{i\nu'-\xi_{\mathbf{p}}} \approx{} -i\frac{N(0)}{2}\int\frac{d\nu' d\theta}{(2\pi)^{2}} \frac{\text{sgn}(\nu'+\nu)-\text{sgn}(\nu')}{i\nu-v_{F}q\cos\theta}.
\end{align}
After a Wick rotation, it can be written as
\begin{align}
    \Pi_0(\Omega,\mathbf{q})\approx{}&N(0)\int\frac{d\omega d\theta}{(2\pi)^2}\frac{f_{0}(\omega+\Omega)-f_{0}(\omega)}{\Omega-v_{F}q\cos\theta},\label{eq:bare_susceptibility_energy_first}
\end{align}
which is the same as Eq.~\eqref{eq:non-int_den_correl}. 

It is known that the evaluation of the bare particle-hole bubble depends on whether the integral on $\xi_p$ or $\omega$ is performed first~\cite{mross2010controlled}. Indeed, if the integral over frequency is performed first, it yields
\begin{align}
\Pi_{0}(i\nu,\mathbf{q})={}&-\int\frac{d\nu'}{2\pi}\frac{d^{2}p}{(2\pi)^{2}} \frac{1}{i\nu'+i\nu-\xi_{\mathbf{p}+\mathbf{q}}}\frac{1}{i\nu'-\xi_{\mathbf{p}}}\notag\\
\approx{}&N(0)\int\frac{d\nu'}{2\pi}\frac{d\theta d\xi_{p}}{(2\pi)^{2}}\frac{1}{i\nu-v_{F}q\cos\theta}\left[\frac{1}{i\nu'+i\nu-\xi_{p}-v_{F}q\cos\theta}-\frac{1}{i\nu'-\xi_{p}}\right]\notag\\
={}&N(0)\int\frac{d\theta d\xi_{p}}{(2\pi)^{2}}\frac{f_{0}(\xi_{p}+v_{F}q\cos\theta)-f_{0}(\xi_{p})}{i\nu-v_{F}q\cos\theta}\notag\\
={}&-\frac{N(0)}{2\pi}\int\frac{d\theta}{2\pi}\frac{v_{F}q\cos\theta}{i\nu-v_{F}q\cos\theta},
\end{align}
which, after the Wick rotation, can be written as
\begin{align}
\Pi_{0}(\Omega,\mathbf{q})={}&\frac{-N(0)}{2\pi}\int\frac{d\theta}{2\pi}\frac{v_{F}q\cos\theta}{\Omega-v_{F}q\cos\theta}
={} \frac{N(0)}{2\pi}\left[1-\int\frac{d\theta}{2\pi}\frac{\Omega}{\Omega-v_{F}q\cos\theta}\right]=N(0)\left[\frac{1}{2\pi}+\int\frac{d\omega d\theta}{(2\pi)^{2}}\frac{f_{0}(\omega+\Omega)-f_{0}(\omega)}{\Omega-v_{F}q\cos\theta}\right].\label{eq:bare_susceptibility_frequency_first}
\end{align}
This integral has a difference in its real part from Eq.~\eqref{eq:bare_susceptibility_energy_first}. To understand the origin of this difference, we recall that the final result of two integrals is independent of the choice of integration order only if the double integral is absolutely convergent. In fact, we know that the bare susceptibility where the integral over frequency first yields the correct physical results since it recovers the known limit $\Pi(\Omega=0,\mathbf{q}\to 0) = N(0)/(2\pi)$. In contrast, the result with the integral over energy performed first yields $\Pi(\Omega=0,\mathbf{q}\to 0) = 0$. Thus, a calculation faithful to the original microscopic theory requires imposing a cutoff on the fermion energy and sending it to infinity only after taking the frequency cut-off to infinity as follows
\begin{align}
\Pi_{0}(i\nu,\mathbf{q})={}&N(0)\int\frac{d\nu'}{2\pi}\frac{d\xi_{\mathbf{p}}d\theta}{(2\pi)^{2}}\frac{1}{i\nu-v_{F}q\cos\theta}\left[\frac{1}{i\nu'+i\nu-\xi_{\mathbf{p}}-v_{F}q\cos\theta}-\frac{1}{i\nu'-\xi_{\mathbf{p}}}\right]\notag\\
={}&-N(0)\int\frac{d\nu'}{2\pi}\frac{d\theta}{(2\pi)^{2}}\frac{1}{i\nu-v_{F}q\cos\theta}\int_{-\Lambda}^{\Lambda}d\xi_{\mathbf{p}}\left[\frac{1}{\xi_{\mathbf{p}}+v_{F}q\cos\theta-i\nu'-i\nu}-\frac{1}{\xi_{\mathbf{p}}-i\nu'}\right]\notag\\
={}&-N(0)\int\frac{d\nu'}{2\pi}\frac{d\theta}{(2\pi)^{2}}\frac{1}{i\nu-v_{F}q\cos\theta}\left[\ln\left(\frac{\Lambda+v_{F}q\cos\theta-i\nu'-i\nu}{-\Lambda+v_{F}q\cos\theta-i\nu'-i\nu}\right)-\ln\left(\frac{\Lambda-i\nu'}{-\Lambda-i\nu'}\right)\right]\notag\\
\approx{}&-N(0)\int\frac{d\nu'}{2\pi}\frac{d\theta}{(2\pi)^{2}}\frac{1}{i\nu-v_{F}q\cos\theta}\left[\ln\left(\frac{\Lambda-i\nu'-i\nu}{-\Lambda-i\nu'-i\nu}\right)+\frac{2\Lambda(v_{F}q\cos\theta)}{\Lambda^{2}+(\omega_{n}+\nu)^{2}}-\ln\left(\frac{\Lambda-i\nu'}{-\Lambda-i\nu'}\right)\right]\notag\\
\approx{}&-N(0)\int\frac{d\theta}{(2\pi)^{2}}\frac{v_{F}q\cos\theta}{i\nu-v_{F}q\cos\theta}.
\end{align}
After the Wick rotation, it can be written as
\begin{align}
\Pi_{0}(\Omega,\mathbf{q})={}&-N(0)\int\frac{d\theta}{(2\pi)^{2}}\frac{v_{F}q\cos\theta}{\Omega-v_{F}q\cos\theta} ={} \frac{N(0)}{2\pi}\left(1-\frac{\Omega}{\sqrt{\Omega^{2}-v_{F}^{2}q^{2}}}\right),
\end{align}
which restores the well-known expression for the particle-hole bubble in two dimensions~\cite{Chubukov2003}. The above discussion highlights the dependence of the bare charge susceptibility on the choice of regularization. However, universal low-energy singularities are expected to be insensitive to the choice of regularization~\cite{mross2010controlled}. Accordingly, we shall not be concerned with using the microscopically accurate regularization scheme in this work and will simply perform the integral over the momentum perpendicular to the Fermi surface first. Any discrepancy can be cured by following the prescriptions described above.

\subsection{Self-energy corrections} \label{app:diagrammatic_calculation_charge_response_self_energy}

The self-energy corrections in Fig.~\ref{fig:Pi1} and~\ref{fig:Pi2} are given in the equilibrium theory for small momentum transfer $\mathbf{q}$ by
\begin{align}
\Pi_{1a}(i\nu,\mathbf{q})={}&-\int\frac{d\nu'}{2\pi}\frac{d^{2}p}{(2\pi)^{2}} \frac{\Sigma^{R}_{0}(i\nu'+i\nu)}{(i\nu'+i\nu-\xi_{\mathbf{p}+\mathbf{q}})^{2}(i\nu'-\xi_{\mathbf{p}})}
-\int\frac{d\nu'}{2\pi}\frac{d^{2}p}{(2\pi)^{2}} \frac{\Sigma^{R}_{0}(i\nu')}{(i\nu'+i\nu-\xi_{\mathbf{p}+\mathbf{q}})(i\nu'-\xi_{\mathbf{p}})^{2}}\notag\\
\approx{}&-\frac{iN(0)}{2}\int\frac{d\nu' d\theta}{(2\pi)^{2}}\frac{(\text{sgn}(\nu'+\nu)-\text{sgn}(\nu'))}{(i\nu-v_{F}q\cos\theta)^{2}}(\Sigma^{R}_{0}(i\nu'+i\nu)-\Sigma^{R}_{0}(i\nu')).
\end{align}
We here use imaginary time (or Euclidean space-time). After the Wick rotation back to real frequencies, the contribution is
\begin{align}   
\Pi_{1a}(\Omega,\mathbf{q})\approx{}&N(0)\int\frac{d\omega d\theta}{(2\pi)^{2}}\frac{(f_{0}(\omega+\Omega)-f_{0}(\omega))}{(\Omega-v_{F}q\cos\theta)^{2}}(\Sigma^{R}_{0}(\omega+\Omega)-\Sigma^{R}_{0}(\omega)).
\end{align}
This is the same as $\langle\delta \hat{\rho} \delta \hat{\rho}\rangle_{1 a}$ in Eq.~\eqref{eq:perturb_a_EFT}.

\subsection{Maki-Thompson diagram}\label{app:diagrammatic_calculation_charge_response_maki_Thompson}

The first-order vertex correction in Fig.~\ref{fig:Pi3}, also known as the Maki-Thompson diagram~\cite{maki1968critical, thompson1970microwave, thompson1971influence}, is
\begin{align}
\Pi_{1b}(i\nu,\mathbf{q})={}&-\int\frac{d\nu_{1}}{2\pi}\frac{d^{2}p_{1}}{(2\pi)^{2}}\int\frac{d\nu_{2}}{2\pi}\frac{d^{2}p_{2}}{(2\pi)^{2}}\frac{D_{1}(i\nu_{2}-i\nu_{1},\theta_{2}-\theta_{1})}{(i\nu_{1}+i\nu-\xi_{\mathbf{p}_{1}+\mathbf{q}})(i\nu_{1}-\xi_{\mathbf{p}_{1}})(i\nu_{2}+i\nu-\xi_{\mathbf{p}_{2}+\mathbf{q}})(i\nu_{2}-\xi_{\mathbf{p}_{2}})}\notag\\
\approx{}&\frac{N(0)^{2}}{4}\int\frac{d\nu_{1}d\theta_{1}}{(2\pi)^{2}}\int\frac{d\nu_{2}d\theta_{2}}{(2\pi)^{2}}\frac{(\text{sgn}(\nu_{1}+\nu)-\text{sgn}(\nu_{1}))(\text{sgn}(\nu_{2}+\nu)-\text{sgn}(\nu_{2}))}{(i\nu-v_{F}q\cos\theta_{1})(i\nu-v_{F}q\cos\theta_{2})}D_{1}(i\nu_{2}-i\nu_{1},\theta_{2}-\theta_{1}),
\end{align}
which, after Wick rotation to real frequency
\begin{align}
    \Pi_{1b}(\Omega,\mathbf{q})\approx{}&-N(0)^{2}\int\frac{d\omega_{1}d\theta_{1}}{(2\pi)^{2}}\int\frac{d\omega_{2}d\theta_{2}}{(2\pi)^{2}}\frac{(f_{0}(\omega_{1}+\Omega)-f_{0}(\omega_{1}))(f_{0}(\omega_{2}+\Omega)-f_{0}(\omega_{2}))}{(\Omega-v_{F}q\cos\theta_{1})(\Omega-v_{F}q\cos\theta_{2})}D_{1}(\omega_{2}-\omega_{1},\theta_{2}-\theta_{1}),
\end{align}
is seen to correspond to $\langle\delta \hat{\rho} \delta \hat{\rho}\rangle_{1 b}$ in Eq.~\eqref{eq:perturb_b_EFT}.

\section{Off-diagonal contribution to the charge response at all loop orders}\label{app:off_diagoanl_AL}

When the quadratic action is expanded at all loop orders, the off-diagonal part of the quadratic action now depends on all $D_n$ parameters as in Eq.~\eqref{eq:general_disc_quad_int}. The first-order contribution to the charge response coming from the off-diagonal interaction vertex containing $D_n$ (i.e., $V_{\text{off},n}$) is given in Eq.~\eqref{eq:charge_response_off_n_loop_order}. We now argue that, in the theory of an interacting Fermi surface with a Yukawa coupling to bosons, the charge response coming from the off-diagonal part of the quadratic action $\expval{\delta\hat{\rho} \delta\hat{\rho}}^{(1)}_{\text{off,}n}$ corresponds to vertex corrections of the bare particle-hole bubble. To see this, it can first be noted that the term involving Fermi-Dirac distributions in Eq~\eqref{eq:charge_response_off_n_loop_order}:
\begin{align}
    N(0)^2\int_{\theta_1 \omega_1 \theta_2 \omega_2}\frac{ F_{0}(\omega_1,\Omega) F_{0}(\omega_2,\Omega) }{ (\Omega - v_{F}(\mathbf{n}_{\theta_1}\cdot\mathbf{q})) (\Omega - v_{F}(\mathbf{n}_{\theta_2}\cdot\mathbf{q}))}
\end{align}
can be obtained in the Fermionic theory by considering the two fermionic Green's functions connecting to the incoming external line and the two Green's functions connecting to the outgoing external line. To see this, consider the following generic vertex correction (of the Fermi surface theory) shown in Fig.~\ref{fig:generic_vertex_correction}, 
\begin{figure}
\center
\includegraphics[width=0.3\linewidth]{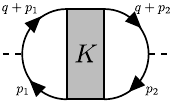}
\caption{Generic vertex correction of the charge response of Fermi surface with a Yukawa coupling to bosonic modes. The full and dotted lines represent fermionic propagators and external vertices, respectively. \label{fig:generic_vertex_correction}}
\end{figure}
where $K$ is a function of the momenta and Matsubara frequencies involved in the four Green's function connecting to the external line (i.e., ($i\nu,\mathbf{q}$), ($i\nu_1,\mathbf{p}_1$), and ($i\nu_2,\mathbf{p}_2$)), that can be obtained by performing integrals over internal momenta and frequencies. As a simple example, we have 
\begin{align}
    K(i\nu,i\nu_1,i\nu_2,\mathbf{q},\mathbf{p}_1,\mathbf{p}_2) = D_1(i\nu_1-i\nu_2, \mathbf{p}_1 - \mathbf{p}_2)
\end{align}
for the Maki-Thomson diagram. 

Evaluating this generic diagram yields expressions of the form
\begin{align}
    I(i\nu,\mathbf{q}) 
    ={}&-\int\frac{d\nu_{1}}{2\pi}\frac{d^{2}p_{1}}{(2\pi)^{2}}\int\frac{d\nu_{2}}{2\pi}\frac{d^{2}p_{2}}{(2\pi)^{2}} \frac{K(i\nu,i\nu_1,i\nu_2,\mathbf{q},\mathbf{p}_1,\mathbf{p}_2)}{(i\nu_{1}+i\nu-\xi_{\mathbf{p}_{1}+\mathbf{q}})(i\nu_{1}-\xi_{\mathbf{p}_{1}})(i\nu_{2}+i\nu-\xi_{\mathbf{p}_{2}+\mathbf{q}})(i\nu_{2}-\xi_{\mathbf{p}_{2}})}\notag\\
    \approx{}&\frac{N(0)^{2}}{4}\int\frac{d\nu_{1}d\theta_{1}d\nu_{2}d\theta_{2}}{(2\pi)^{4}} \frac{ (\text{sgn}(\nu_{1}+\nu)-\text{sgn}(\nu_{1}))(\text{sgn}(\nu_{2}+\nu)-\text{sgn}(\nu_{2}))}{(i\nu-v_{F}q\cos\theta_{1})(i\nu-v_{F}q\cos\theta_{2})}K(i\nu,i\nu_1,\mathbf{q},p_F \hat{\theta}_1,p_F \hat{\theta}_2)
\end{align}
which after Wick rotation to real frequency is
\begin{align}
    I(\Omega,\mathbf{q})\approx{}&-N(0)^{2}\int\frac{d\omega_{1}d\theta_{1}d\omega_{2}d\theta_{2}}{(2\pi)^{4}} \frac{(F_{0}(\omega_1,\Omega) F_{0}(\omega_2,\Omega)}{(\Omega-v_{F}q\cos\theta_{1})(\Omega-v_{F}q\cos\theta_{2})}  K(\Omega,\omega_1,\omega_2,\mathbf{q},p_F \hat{\theta}_1,p_F \hat{\theta}_2).
\end{align}
The momenta $\mathbf{p}_1$ and $\mathbf{p}_2$ were fixed on the Fermi surface by the PK reduction procedure. By direct comparison with Eq.~\eqref{eq:charge_response_off_n_loop_order}, we then see that the contribution in the bozonized theory $\expval{\delta\hat{\rho}\delta\hat{\rho}}_{\text{off,}n}^{(1)}$ corresponds (in the Fermi surface theory) to a vertex correction to the charge response with the identification
\begin{align}\label{eq:identification_K}
    &K(\Omega,\omega_1,\omega_2,\mathbf{q},p_F \hat{\theta}_1,p_F \hat{\theta}_2) = n \int_{\omega'_1 \theta'_1}\ldots\int_{\omega_{n-1}'\theta_{n-1}'}  \left(\prod_{i=1}^{n-1}g_{0}(\omega_{i}')\right) D_{n}(\omega_{1}'-\omega_1,\theta_{1}'-\theta_1;\ldots;\omega_2-\omega_1,\theta_2-\theta_1) 
\end{align}

Recalling that 
\begin{align}
    \Sigma&^{R(n)}[g](\omega,\theta)=\int_{\omega_{1}\theta_{1}}\ldots\int_{\omega_{n}\theta_{n}}D_{n}(\omega_{1}-\omega,\theta_{1}-\theta;\ldots;\omega_{n}-\omega,\theta_{n}-\theta) \prod_{i=1}^{n} g(t,\mathbf{x};\omega_{i},\theta_{i}),
\end{align}
we see that the right-hand side of Eq.~\eqref{eq:identification_K} is
\begin{align}
     \fdv{\Sigma^{R(n)}[g_0](\omega_1,\theta_1)}{g_0(\omega_2,\theta_2)} &= n \int_{\omega'_1 \theta'_1}\ldots\int_{\omega_{n-1}'\theta_{n-1}'}  \left(\prod_{i=1}^{n-1}g_{0}(\omega_{i}')\right) D_{n}(\omega_{1}'-\omega_1,\theta_{1}'-\theta_1;\ldots;\omega_2-\omega_1,\theta_2-\theta_1).
\end{align}
Thus, the diagrammatic corrections to the charge response produced by the off-diagonal vertices containing $D_n$ (i.e., $\expval{\delta\hat{\rho}\delta\hat{\rho}}_{\text{off,}n}^{(1)}$) corresponds heuristically to diagrams in the fermionic theory that are obtained by removing a fermionic line to the $n$-loop self-energy $\Sigma^{R(n)}_0$ and associating the resulting diagram with the vertex correction $K$. 

As a simple example, let's consider the contribution $\expval{\delta\hat{\rho}\delta\hat{\rho}}_{\text{off,}1}^{(1)}$. To see the corresponding diagrammatic contribution in the fermion language, we take the one-loop self-energy $\Sigma^{R(1)}_0$ (i.e., the fock self-energy diagram) and identify $K$ with the result of removing one fermionic line. Following this procedure, we get the Maki-Thomson diagram as illustrated in Fig.~\ref{fig:vertex_correction_procedure_MT}. 
\begin{figure}
\center
\includegraphics[width=0.75\linewidth]{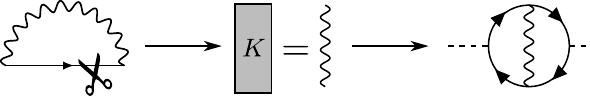}
\caption{Heuristic procedure to associate $\expval{\delta\hat{\rho}\delta\hat{\rho}}_{\text{off,}1}^{(1)}$ with the Maki-Thomson diagram. A fermionic propagator is first removed from  $\Sigma^{R(1)}_0$ (left), and the resulting diagram is then associated with the vertex correction $K$ (right). \label{fig:vertex_correction_procedure_MT}}
\end{figure}

Let's now consider the contribution to the charge response $\expval{\delta\hat{\rho}\delta\hat{\rho}}_{\text{off,}3}^{(1)}$ that contains $D_3$. If the procedure outlined above is followed and we remove a fermionic propagator from $\Sigma^{R(3)}_0$ to generate a vertex correction, Fig.~\ref{fig:vertex_correction_procedure_AL} shows that, for instance, Aslamazov-Larkin diagrams can be obtained. Showing an exact correspondence would involve considering a specific theory (i.e., a specific form of $D_1$ as well as a specific relation between $D_{n}$ with $n>1$ and $D_1$).

\begin{figure}
\center
\includegraphics[width=0.85\linewidth]{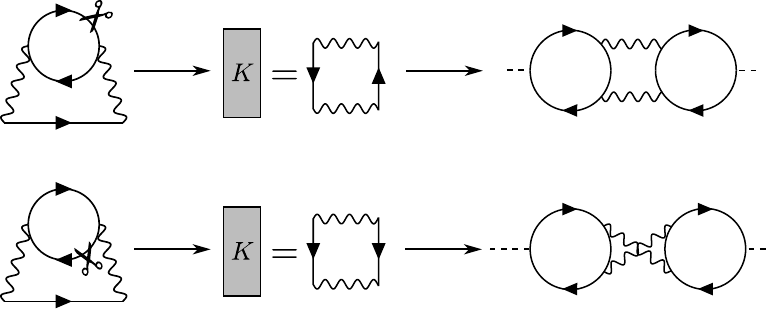}
\caption{Heuristic procedure to show how $\expval{\delta\hat{\rho}\delta\hat{\rho}}_{\text{off,}n}^{(1)}$ contains the Aslamazov-Larkin diagrams. A fermionic propagator is first removed from  $\Sigma^{R(3)}_0$ (left), and the resulting diagram is then associated with the vertex correction $K$ (right). \label{fig:vertex_correction_procedure_AL}}
\end{figure}

\section{Comments on non-linear corrections in the general case}\label{app:general_higher}

Here, we will discuss the higher-order expansion of the action in terms of $\phi$, where we consider the full functional expansion of the self-energy. The general effective action Eq.~\eqref{eq:action} can be expanded as
\begin{align}
S={}&\braket{\tilde{g},\omega-\xi_{\mathbf{p}}-\sum_{n=1}\tfrac{1}{n+1}\Sigma^{R(n)}[\tilde{g}]}\notag\\
={}&\braket{\tilde{g}_{0}+\Delta_{\phi}^{(1)}\tilde{g}+\Delta_{\phi}^{(2)}\tilde{g}+\ldots,\omega-\xi_{\mathbf{p}}-\sum_{n=1}\tfrac{1}{n+1}(\Sigma^{R(n)}[\tilde{g}_{0}]+\Delta_{\phi}^{(1)}\Sigma^{R(n)}+\Delta_{\phi}^{(2)}\Sigma^{R(n)}+\ldots)}\notag\\
={}&\braket{\tilde{g}_{0},\omega-\xi_{\mathbf{p}}-\sum_{n=1}\tfrac{1}{n+1}\Sigma^{R(n)}[\tilde{g}_{0}]}\notag\\
&+\braket{\Delta_{\phi}^{(1)}\tilde{g},\omega-\xi_{\mathbf{p}}-\Sigma^{R}[\tilde{g}_{0}]}\notag\\
&+\braket{\Delta_{\phi}^{(2)}\tilde{g},\omega-\xi_{\mathbf{p}}-\Sigma^{R}[\tilde{g}_{0}]}-\sum_{n=1}\frac{n}{2}\braket{\Delta_{\phi}^{(1)}\tilde{g},\bar{\Delta}_{\phi}^{(1)}\Sigma^{R(n)}}\notag\\
&+\ldots.
\label{eq:general_effective_expand}
\end{align}
From Eq.~\eqref{eq:general_effective_expand}, we can obtain the general form of $m$-th order action in $\phi$ given by
\begin{align}
    S^{(m)}={}&\braket{\Delta_{\phi}^{(m)}\tilde{g},\omega-\xi_{\mathbf{p}}-\Sigma^{R}[\tilde{g}_{0}]}-\sum_{n=1}\sum_{\{m_{i}\}}C_{n}(\{m_{i}\})\braket{\Delta_{\phi}^{(m_{1})}\tilde{g},\bar{\Delta}_{\phi}^{(m_{2},\ldots,m_{j})}\Sigma^{R(n)}},\label{eq:general_expansion_form}
\end{align}
where $\{m_{i}\}=(m_{1},m_{2}\ldots,m_{j})$ is a partition for $m$. So we perform the summation on all the possible partitions of $m$, except $\{m_{i}\}=m$ because it is already included in the first term in Eq.~\eqref{eq:general_expansion_form}. The coefficient $C_{n}(\{m_{i}\})$ is the number of possibilities for choosing $\{m_{i}\}$ combinations divided by $(n+1)$, which is defined by $C_{n}(\{m_{i}\})=\frac{n!}{(n+1-j)!\prod_{i=1}n_{i}!}$, and $n_{i}$ is the number of $i$ in the partition and $j$ is the size of partition $\{m_{i}\}$. For example, in the case of $\int_{t\omega}\braket{\Delta_{\phi}^{(3)},\bar{\Delta}_{\phi}^{(3,2,2,1)}\Sigma^{R(n)}}$, which is one of the $\phi^{11}$ order contributions, since $n_{3}=2$, $n_{2}=2$, $n_{1}=1$, we have $C_{n}(3,3,2,2,1)=\frac{n!}{2!2!1!(n-4)!}=n(n-1)(n-2)(n-3)/4$.

As another simple example, the cubic action in $\phi$ is given by
\begin{align}
    S^{(3)}={}&\braket{\Delta_{\phi}^{(3)}\tilde{g},\omega-\xi_{p}-\Sigma^{R}[\tilde{g}_{0}]}-\sum_{n=1}n\braket{\Delta_{\phi}^{(2)}\tilde{g},\bar{\Delta}_{\phi}^{(1)}\Sigma^{R(n)}}
-\sum_{n=1}\frac{n(n-1)}{6}\braket{\Delta_{\phi}^{(1)}\tilde{g},\bar{\Delta}_{\phi}^{(1,1)}\Sigma^{R(n)}}.\label{eq:general_cubic}
\end{align}
The last term in Eq.~\eqref{eq:general_cubic} is completely new and not found in the one-loop order approximation (Eqs.~\eqref{eq:cont_cubic_bare} and \eqref{eq:cont_cubic_int}).
The coefficients of the second and last terms are $C_{n}(2,1)=\frac{n!}{1!1!(n-1)!}=n$ and $C_{n}(1,1,1)=\frac{n!}{3!(n-2)!}=n(n-1)/6$, respectively. That is because $n_{2}=1$, $n_{1}=1$, and $j=2$ for the second term, whereas $n_{1}=3$ and $j=3$ for the last term. For the first term, $C_{n}(3)=\frac{n!}{1!n!}=1$ since $n_{3}=1$.

\end{widetext}

%

\end{document}